\documentclass[a4paper,twoside]{article}
\newcommand{\affil}[1]{$^{\rm #1}$}
\usepackage[authoryear]{natbib}
\bibpunct{(}{)}{;}{a}{}{,}
\usepackage{graphicx}
\usepackage{amssymb}
\usepackage{amsmath}
\usepackage{pstricks}
\date{} 
\newcommand{\ud}{\mathrm{d}} 
\newcommand{\Mo}{\rm{M}_\odot}
\newcommand{\Lo}{\rm{L}_\odot}
\newcommand{\Ro}{\rm{R}_\odot}

\def\NB6{\texttt{NBODY6}}
\def\STARS{\texttt{STARS}}
\def\SSE{\texttt{SSE}}
\def\name{\mathcal{N}}
\title{\large\bf\flushleft $N$-body Simulations with Live Stellar Evolution}
\author{\parbox{\textwidth}{\flushleft
\vspace{-0.5cm}
{\it Ross P. Church\affil{A,B,D}, Christopher A. Tout\affil{B}, and Jarrod R.
Hurley\affil{A,C}}\\
\vspace{0.4cm}
{\small \affil{A}\,Centre for Stellar and Planetary Astrophysics, School of Mathematical Sciences, Monash University, Vic. 3800}\\
{\small \affil{B}\,Institute of Astronomy, The Observatories, Madingley Road, Cambridge CB3 0HA, England}\\
{\small \affil{C}\,Centre for Astrophysics and Supercomputing,
Swinburne University of Technology,
Hawthorn Vic. 3122}\\
{\small \affil{D}\,Email: ross.church@sci.monash.edu.au}}}
\begin{document}
\twocolumn[
\begin{changemargin}{.8cm}{.5cm}
\begin{minipage}{.9\textwidth}
\vspace{-1cm}
\maketitle
%
%

\small{\bf Abstract:}
An $N$-body code containing live stellar evolution through combination of the
software packages \NB6 and \STARS\ is presented.  Operational details of the two
codes are outlined and the changes that have been made to combine them
discussed.  We have computed the evolution of clusters of $10^4$ stars using the
combined code and we compare the results with those obtained using \NB6 and the
synthetic stellar evolution code \SSE.  We find that, providing the physics
package within \STARS\ is set up correctly to match the parameters of the models
used to construct \SSE, the results are very similar.  This provides a good
indication that the new code is working well.  We also demonstrate how this
physics can be changed simply in the new code with convective overshooting as an
example.  Similar changes in \SSE{} would require considerable reworking of the
model fits.  We conclude by outlining proposed future development of the code to
include more complete models of single stars and binary star systems.  


\medskip
\medskip
\end{minipage}
\end{changemargin}
]
\small

\section{Introduction} 
The gravitational $N$-body problem -- the motion of a set of massive bodies
under mutual self-gravity -- is one of the oldest in computational astronomy.
The first simulations were carried out by \citet{1941ApJ....94..385H} who used
special-purpose hardware.  As computing power has increased, particularly
with the introduction of the GRAPE hardware~\citep{1998sssp.book.....M} and
computational algorithms have improved it has become possible to include greater
numbers of particles and more sophisticated physics in order to make
more accurate models of star clusters, galaxies and the Universe itself.
To study the evolution of galactic and globular open clusters we must integrate
the motion of each particle separately, considering the force contribution from
each of the other particles.  Because of the high densities in the cores of
stellar clusters particles can approach very close to one another and we must
include special procedures to prevent such encounters introducing unacceptable
errors.

To make physically realistic models of stellar clusters we cannot avoid
considering the evolution of the stars themselves.  Mass loss from stars changes
the cluster's potential and hence affects the motion of its stars.  Furthermore
the finite radius of a star means that if it approaches close to another they
may interact.  Tidal dissipation may cause a bound binary system to form, within
which matter can be transferred from one star to another, and stars that come
sufficiently close may merge.  To predict the evolution of the mass and radius
of a star with time it is necessary to use a stellar evolution code.  Stellar
evolution is another area of computational astronomy with a considerable
heritage.  A procedure for evaluating numerical models of stars by use of an
electronic computer was outlined by \citet{1956MNRAS.116..515H}.  Over the years
such simulations have grown in detail and accuracy as more physics has been
added and more powerful computers have become available for the integration of
the equations.  In the past however the computational cost of these calculations
has prohibited including them in cluster simulations.  

The most popular approach to date for including stellar evolution in cluster
models has been the use of synthetic stellar evolution codes.  This involves
analytic fits made to the results of a full code evaluated to simulate the
evolution of a star.  A prime example is the SSE code
of~\citet*{2002MNRAS.329..897H} which was derived from the detailed models of
\citet{1998MNRAS.298..525P}.  The combination of synthetic stellar evolution and
$N$-body dynamics has been used to good effect to model the interplay between
stellar and dynamical evolution in star clusters
\citep{2001MNRAS.321..199P,2003ApJ...589L..25B,2005MNRAS.363..293H}.  However, a
major drawback is that the fitting process is laborious and the result
inflexible, in that it relies on choices made in generating the underlying set
of detailed models.  If the input physics used for the detailed models becomes
out of date in any way then so does the synthetic package and it is non-trivial
to generate a new set of analytic fits.  It is also ill-equipped to deal with
non-equilibrium cases arising from binary mass-transfer and stellar collisions.
The pioneers of the synthetic stellar evolution approach were
\citet*{1989ApJ...347..998E} who preferred this to the main alternative at the
time which was interpolation within a database of detailed models.  It was
decided that such a database would be cumbersome -- especially if a large grid
of models of various mass and metallicity were required -- and create problems
for data storage.  Fortunately this is not a major issue anymore and the
interpolation approach does have its merits.  The ideal is to perform live
stellar evolution in combination with $N$-body dynamics and to also include a
module for dealing with any stellar collisions that arise, for example the
smooth particle hydrodynamics of \citet{2003NewA....8..605S}.

Computing power has increased in recent times to the point where it is possible
to combine a full stellar evolution code with a $N$-body code.  We present here
such a code and compare the results to those obtained with synthetic stellar
evolution.  Section~\ref{sect:codes} contains details of the two codes that we
used and the models that we compare are described in Section~\ref{sect:setup}.
The results are presented in Section~\ref{sect:results} and discussed in
Section~\ref{sect:discuss}.

\section{Description of the code}
\label{sect:codes}
We give a brief description of some of the relevant features of the two codes,
\STARS{} and \NB6, along with the modifications that have been necessary to
combine them.

\subsection{\STARS}
\STARS, the Cambridge Stellar Evolution code, was originally written by 
\citet{1971MNRAS.151..351E} and has been extensively modified since (see
\citealt{1995MNRAS.274..964P} and references therein for a complete description).
\STARS~utilises the method of \citet{1964ApJ...139..306H}.  The equations of
stellar structure are written as implicit finite difference equations on a mesh
and solved by numerical inversion of the resulting matrix.  The mesh has a fixed
number of meshpoints that can move in both mass and radius, and hence is neither
Eulerian nor Lagrangian.  The model is written in terms of the quantities $\log
f$ (a measure of electron degeneracy as described by
\citealt*{1973A&A....23..325E}), temperature, mass, radius, luminosity, the mass
fractions of hydrogen, helium-4, carbon-12, oxygen-16 and neon-20 and a quantity
$Q$ which determines the position of the mesh.  In a converged model the
gradient of $Q$ is equal at all points on the mesh.  It has an ad-hoc functional
form that causes more points to be placed where the temperature, mass, pressure
and radius are varying most quickly.  These are the regions, such as burning
shells and ionisation zones in the atmosphere, that need to be well resolved.
The system of equations is then solved by a relaxation method.  The timestep in
\STARS~is controlled by the user supplied parameter $\Delta$.  After a model has
converged the next timestep is calculated as \begin{equation} \delta \tau_{i+1}
= \delta \tau_i \times \frac{\Delta}{\sum_{jk} |\delta X_{jk}|} \end{equation}
where $\delta X_{jk}$ is the change in variable $j$ at meshpoint $k$.  The sum
is evaluated over all the variables except the luminosity.  A larger
value of $\Delta$ allows the variables to change more in a single timestep and
hence larger timesteps are taken.  Convective overshooting can be included by
means of a modified Schwarzschild criterion controlled by a single parameter
$\delta_{\rm ov}$; details of the implementation are described in
\citet{1997MNRAS.285..696S}.

\subsubsection{Zero-age models}
The models described in this paper were all calculated at solar composition,
$X=0.7$, $Y=0.28$ and $Z=0.02$.  A library of zero-age main-sequence models of
masses between $0.1\,\Mo$ and $100\,\Mo$ has been produced by a multi-stage
process.  We took a stellar model of uniform solar composition and inserted an
artificial energy source to inflate it into a cold, low-density cloud until the
core temperature was below $10^6$\,K, below which no nuclear reactions are
modelled by the code.  We then added and removed mass to produce a pre-main
sequence of models of low density and temperature.  We removed the artificial
energy source and followed the contraction down to the point of minimum radius
to produce a set of models with self-consistent central compositions and
temperatures.  By adding a small amount of mass to the envelope of one of
these models we can construct models of zero-age main-sequence stars with any
mass.

\subsubsection{The helium flash}
We have developed a routine to pseudo-evolve through the helium flash in
low-mass stars.  In stars less massive than about $2.3\,\Mo$ the core is degenerate at the
time of helium ignition so that the increased temperature owing to helium
burning does not cause expansion and thermal runaway occurs
\citep{1962ApJ...136..158S}.  To evolve through the helium flash with
\STARS~requires very small timesteps, which lead to numerical instability in
calculation of the luminosity.  To circumvent these problems we construct
approximate post-flash models with stable core helium burning.  A star of mass
$M\simeq3\,\Mo$ that has evolved successfully through non-degenerate core helium
ignition is taken and matter removed from the envelope until the desired mass is
reached.  The hydrogen burning shell is then allowed to burn outwards with
helium consumption disabled in order to obtain the correct core mass and the
envelope compositions reset to their pre-flash values.  Normal evolution is then
resumed.  Whilst not physically rigorous this process provides models that can
be used to simulate subsequent evolution.

\subsubsection{Mass loss}
We have written a simple procedure to determine the type of a star (main
sequence, red giant, etc.) from the central abundances and hydrogen, helium and
carbon burning luminosities of the star.  Then we automatically choose an
appropriate mass-loss law for the star according to
\begin{itemize}
\item Main sequence -- no mass loss,
\item Red giant branch (RGB) -- Reimers mass loss \citep{1978A&A....70..227K}
\begin{equation}
\frac{\ud M}{\ud t} = -4\times10^{-13}\eta\frac{L}{\Lo}\frac{R}{\Ro}\frac{\Mo}{M}\Mo\rm{yr}^{-1},
\end{equation}
with $\eta=0.4$,
\item Core helium burning -- Reimers mass loss with $\eta=1.0$,
\item Asymptotic giant branch (AGB) -- \citet{1993ApJ...413..641V} mass loss 
\begin{multline}
\frac{\ud M}{\ud t} = -\min\left\{10^{-11.4 +
0.0123P/\rm{days}}\Mo\rm{yr}^{-1},\right.\\
\left.\frac{L/c}{(-13.5+0.056P/\rm{days})\,\rm{km\,s}^{-1}} \right\},
\end{multline}
and
\item White dwarf (WD) -- No mass loss.
\end{itemize}

\subsubsection{Timestep and mesh size}
To reduce model runtime and memory requirements we use a relatively low
resolution (199 meshpoints per model).  Runtime is further decreased by choosing
a comparatively large value of the timestep parameter, $\Delta=5$, which ensures
rapid progress throughout the star's life.    These two choices have the added
advantage of suppressing thermal pulses on the AGB, which are too
computationally demanding and numerically difficult to be included in these
models~\citep{2004MNRAS.352..984S}.

\subsubsection{The post-AGB}
Some extra procedures are necessary to complete the evolution of the stars
from the late AGB to the white dwarf cooling track.  This phase of evolution,
known as the post-AGB, is even more numerically demanding than the AGB phase
that proceeds it.  To prevent numerical convergence issues arising from very
high luminosities, once the stars have entered the superwind phase the rates of
the triple-alpha and CNO reactions are capped at $10^{-12}$ and $10^{-11}$
reactions per baryon per second.  Unmodified, \STARS{} experiences a
cyclical phase of evolution where fierce burning in the shell causes the very
light envelope to be blown off.  The shell extinguishes and then reignites as
the envelope contracts back on to the star.  This is both slow to converge, as
many hundreds of cycles can occur even in the presence of strong mass loss and
frequently causes code non-convergence.  This behaviour is thought to be a
numerical artefact of the \STARS\ code though its cause is unclear
\citep{stancliffe05}.  In order to surpress these cycles, once the
hydrogen-burning shell reaches $0.02\,\Mo$ from the surface of the star the
rate of energy generation from nuclear burning is further reduced such that
\begin{equation}
\frac{\ud E}{\ud t} = \left(\frac{m_{\rm env}}{0.02\,\Mo}\right)^{\!\!3} \frac{\ud E}{\ud t}_{\rm physical},
\end{equation}
where $m_{\rm env}$ is the envelope mass (mass between the shell and surface of
the star).  The rate of consumption of material remains at the physical value,
however, so the overall effect is to make the remaining nuclear burning produce
less energy.

As these procedures are both implemented only after the superwind phase of mass
loss has begun, which truncates the AGB, they have very little effect on the
observable evolution.  The most apparent consequence is that the luminosity on
the post-AGB is reduced by about 0.5\,dex; however, this phase is very
short-lived and hence unlikely to be observed; no post-AGB stars appear in the
HR diagrams plotted Figures~1 to 3.  The principal effect of these two
additional procedures is to render the predicted surface compositions of white
dwarfs unreliable.  These compositions are not trustworthy anyway because we do
not model thermal pulses, which radically alter the compositions of AGB stars.
We have found this procedure sufficient to evolve stars of initial masses up to
$4\,\Mo$ for a few Gyr, the typical lifetime of the clusters under
consideration.

\subsubsection{Binaries}
The only element of binary stellar evolution implemented at present is the
collision of stars.  This is handled in a very simple manner.  A zero-age star
of the same mass as the combined mass of the two colliding stars is generated.
Other interactions within binaries that form dynamically are ignored.

\subsection{\NB6}
\NB6 is a general-purpose full force summation $N$-body dynamics code.  A
general description of the development of \NB6 and its sister codes can be found
in a review \citep{1999PASP..111.1333A} and an exhaustive description of
its algorithmic basis, construction and operation in a recent
book~\citep{Aarseth-Book}.  It uses the \citet{AhmadCohen} neighbour scheme to
reduce the cost of computation for large models, where the forces from particles
lying within some local neighbour radius are updated more frequently than those
from particles lying further away.  An Hermite integrator is used with
individual hierarchical timesteps.  To remove the accuracy and performance
problems associated with integrating perturbed binaries around many
orbits~\citet{KS} regularisation is used.  This involves a change of variables
in four dimensions, with one fictitious dimension added to make the
transformation possible, as well as a time scaling.  For perturbed hierarchical
configurations (triples, quadruples, etc.) chain
regularisation~\citep{1990CeMDA..47..375M} is used.  By default \NB6 uses the
synthetic stellar evolution code \SSE~\citep*{2000MNRAS.315..543H}.

To integrate \STARS~with \NB6 the procedures that involve calls to \SSE\ 
have been re-written and a set of subroutines added to provide an interface
between the two codes.  To interrogate the state of star $\name$ at time $t$ a
routine has been written that extracts the required quantities from the live
stellar evolution models at the time $t$ (the evolution and $N$-body codes have
independent time steps).  As a star evolves a model may occasionally fail to
converge; in that case \STARS{} restarts from the previous time with a reduced
timestep.  Hence it is possible for values at the current timestep to change,
because the restarted model may have different evolution.  To avoid information
provided to the $N$-body code being subsequently revised we ensure that the time
for which the data are requested is between the previous and anteprevious
timestep, i.e. that
\[\tau_{\name i-2} < t < \tau_{\name i-1}\]
where $\tau_{\name i}$ is the latest stellar evolution time for star $\name$.
If this is not the case then the model for star $\name$ is loaded into memory
and evolved for the necessary number of timesteps.  Once this has been done the
radius, mass and luminosity at time $t$ are calculated by linear interpolation
between adjacent stellar models and returned.  A time for the next interrogation
of the stellar model is also provided to \NB6.  It is calculated as the maximum
time in which the radius should not change by more than 2\% or the mass by more
than 1\%.  An arbitrary limit of ten \STARS\ timesteps is also imposed,
preventing \NB6 advancing past any interesting phases of evolution that start
suddenly.

\section{Model setup}
\label{sect:setup}
To investigate the differences between models with synthetic and full stellar
evolution five cluster models, each containing $10^4$ stars, were evolved with
our combined code.  For the purposes of comparison we also evolved two clusters
with the same initial conditions but with synthetic stellar evolution.
Subsequent analysis of the models produced caused us to run another model with
full stellar evolution and convective overshooting, with the parameter
$\delta_{\rm ov}=0.12$ as in the construction of the \SSE\ models.  All stellar
models started from the zero-age main sequence and had metallicity $Z=0.02$.

The stars are all taken to be single; there are no primordial binaries.
The masses of the stars are distributed at random from a
\citet*{1993MNRAS.262..545K} initial mass function (IMF), obtained from the
generating function
\begin{equation}
m(X) = 0.08 + \frac{\gamma_1 X^{\gamma_2} + \gamma_3
X^{\gamma_4}}{(1-X)^{0.58}},
\end{equation}
where $X$ is a number randomly chosen from the uniform distribution in the range
[0,1], $\gamma_1=0.19$, $\gamma_2=1.55$, $\gamma_3=0.05$ and $\gamma_4=0.6$.
The initial positions of the particles are distributed according to the Plummer
distribution,
\begin{equation}
\rho(\mathbf{r}) = \frac{3M}{4\pi r^3_0} \frac{1}{[1+(r/r_0)^2]^{5/2}},
\end{equation}
where $M$ is the total cluster mass and $r_0$ is a scaling factor related
through integration to the half-mass radius $r_h$ by $r_h\simeq 1.3\,r_0$.
Following the standard approach for $N$-body units and initial conditions we set
$M = 1$ and $r_0 = 1$.  The distance scaling of the simulation is then
determined by specifying the physical extent of the $N$-body length unit.  We
choose this to be $2\,\rm{pc}$ for all these simulations which, along with the
IMF and imposition of initial virial equilibrium, defines the scaling.  More
details can be found in~\citet{Aarseth-Book}.  A standard tidal field based on
Oort's constants \citep{1927BAN.....3..275O} is applied which places the cluster
at Sun's position in the Galaxy.  We take Oort's constants to be $A = 14.4{\rm
km\,s^{-1}\,kpc^{-1}}$ and $B = -12.0{\rm km\,s^{-1}\,kpc^{-1}}$.  The initial
conditions for the five models were identical except for the random number
generator seed which was changed to give a different set of masses, positions
and velocities for the stars.  Hence the models can be considered to be a small,
representative sample of possible $10^4$ star clusters with that set of initial
conditions.  

\section{Results and comparison}
\label{sect:results}
We extracted various different properties from the data output by the
code,  chosen to measure different aspects of the stellar evolution and dynamics
of the clusters in an attempt to determine how much the simulations differ.  We
have considered the quantities as a function of the fraction of stars remaining
instead of time because this is more characteristic of the dynamical state of
evolution of the cluster and hence there is less scatter between the lines.  In
the case of the graph of the time against fraction of stars remaining this is
counterintuitive but aids comparison with the other graphs in the paper.  For
most of the quantities we have also plotted the standard deviation of the values
from the five standard models (full stellar evolution without convective
overshooting).  The points when the models have a certain fraction of stars
remaining are not exactly coincident.  Consequently we have interpolated the
quantities to a standard set of points.  This is a reasonable approach to take
for quantities that are varying slowly, i.e. on a timescale longer than the
interval between the snapshots of the cluster that the code produces.  This
standard deviation provides a measure of the spread of values owing to the
variation within the population of clusters of the type that we are considering.
Furthermore we have calculated the mean of the five standard models at each of
these points and subtracted it from the models using \SSE\ and \STARS\ with
convective overshooting to get a set of residuals.  By comparing these residuals
with the standard deviations we measure whether the different stellar evolution
prescriptions significantly change the models.

\subsection{HR diagrams}
\begin{figure*}
\begin{center}
\includegraphics[width=\columnwidth]{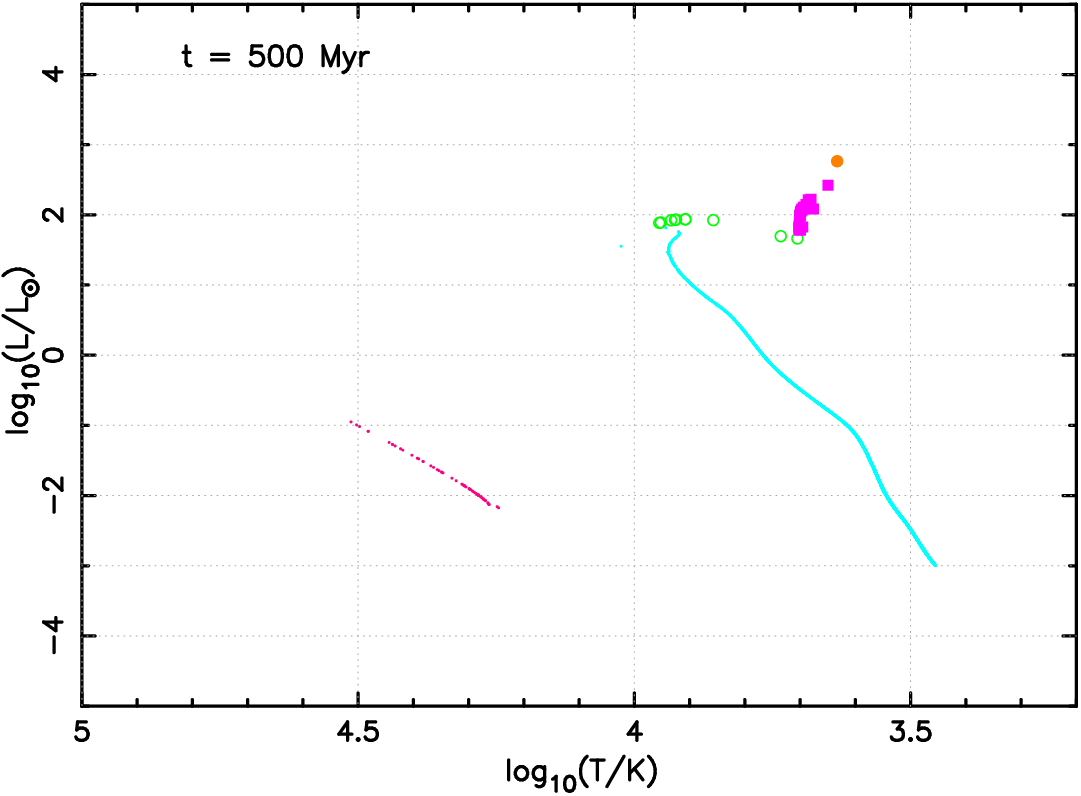}
\includegraphics[width=\columnwidth]{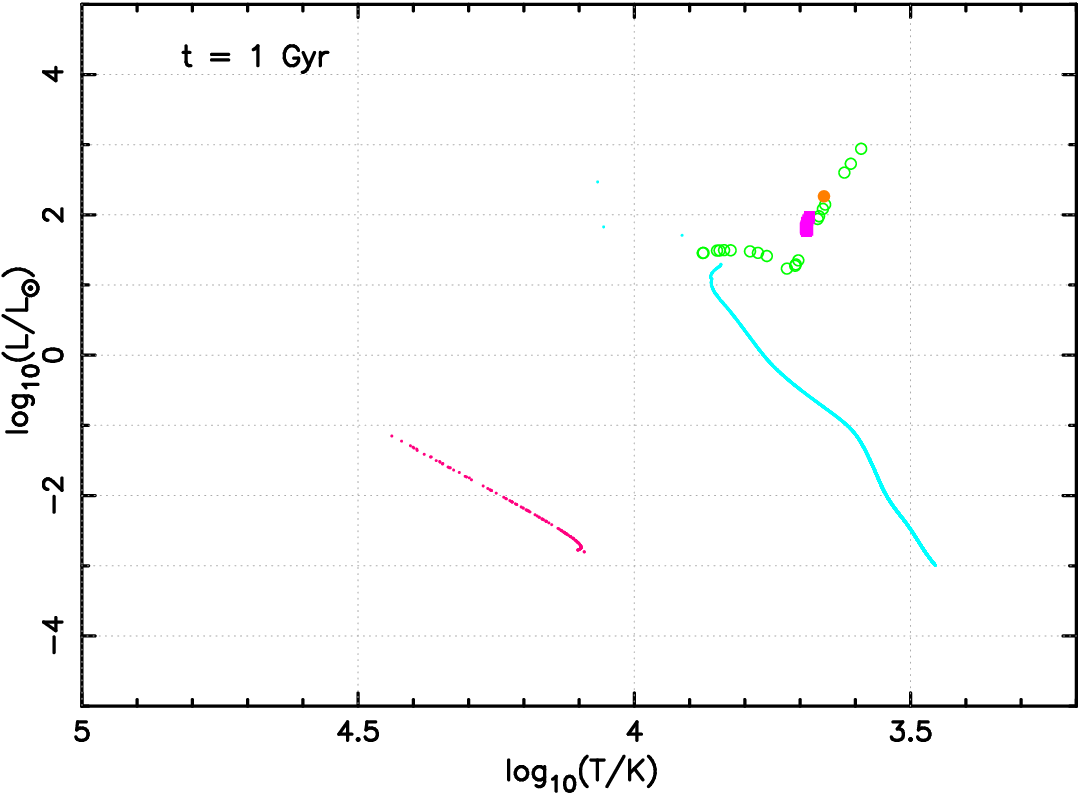}
\includegraphics[width=\columnwidth]{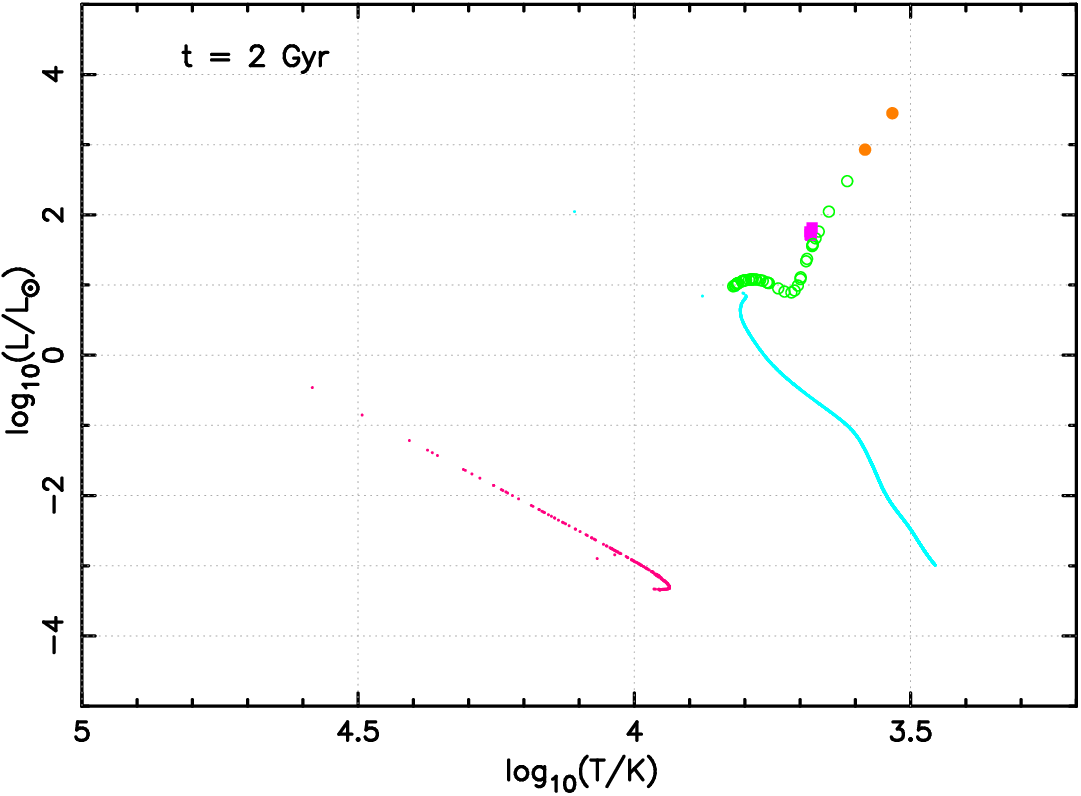}
\includegraphics[width=\columnwidth]{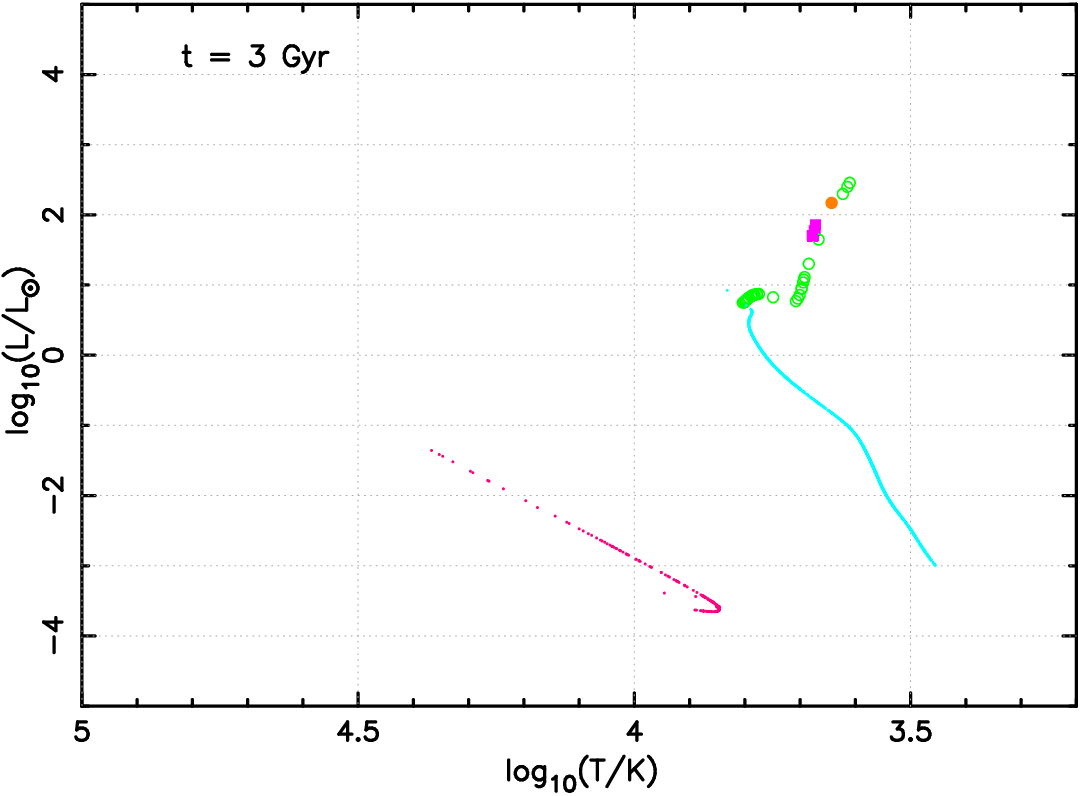}
\includegraphics[width=\columnwidth]{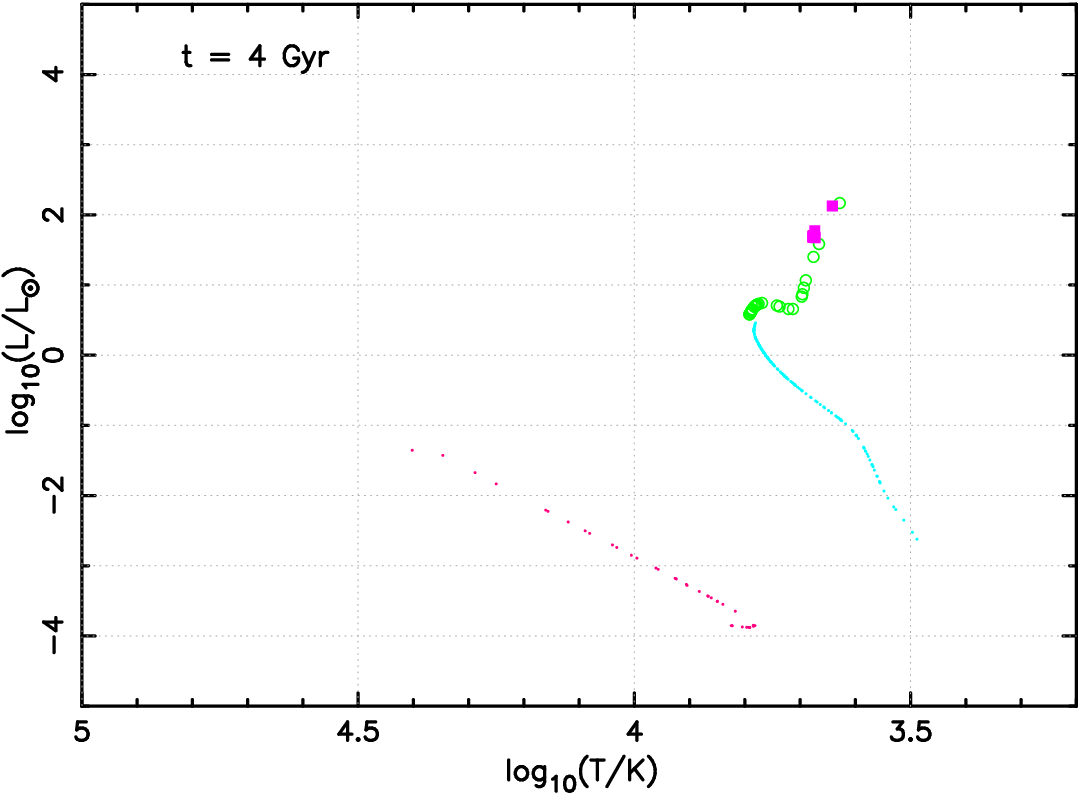}
\caption{HR diagrams for one of the \STARS~models with no convective
overshooting.  Different symbols represent different types of stars.  The thick
line of dots is the main sequence, and the sparser line of dots is the white
dwarf cooling track.  Open circles are Hertzsprung gap and first giant branch
stars, and closed circles AGB stars.  Filled squares are core helium burning
stars.  All binaries are assumed to be fully resolved (i.e. both stars are
plotted separately).  The HR diagrams are plotted at 500\,Myr (top left),
1000\,Myr (top right), 2000\,Myr (centre left), 3000\,Myr (centre right) and
4000\,Myr (bottom).  The small numbers of stars where the model failed to
converge -- mostly stars with masses greater than $4\,\Mo$, formed by collisions
in perturbed binaries that have formed dynamically -- have been removed from the
diagrams.}
\label{fig:HR:STARS}
\end{center}
\end{figure*}

\begin{figure*}
\begin{center}
\includegraphics[width=\columnwidth]{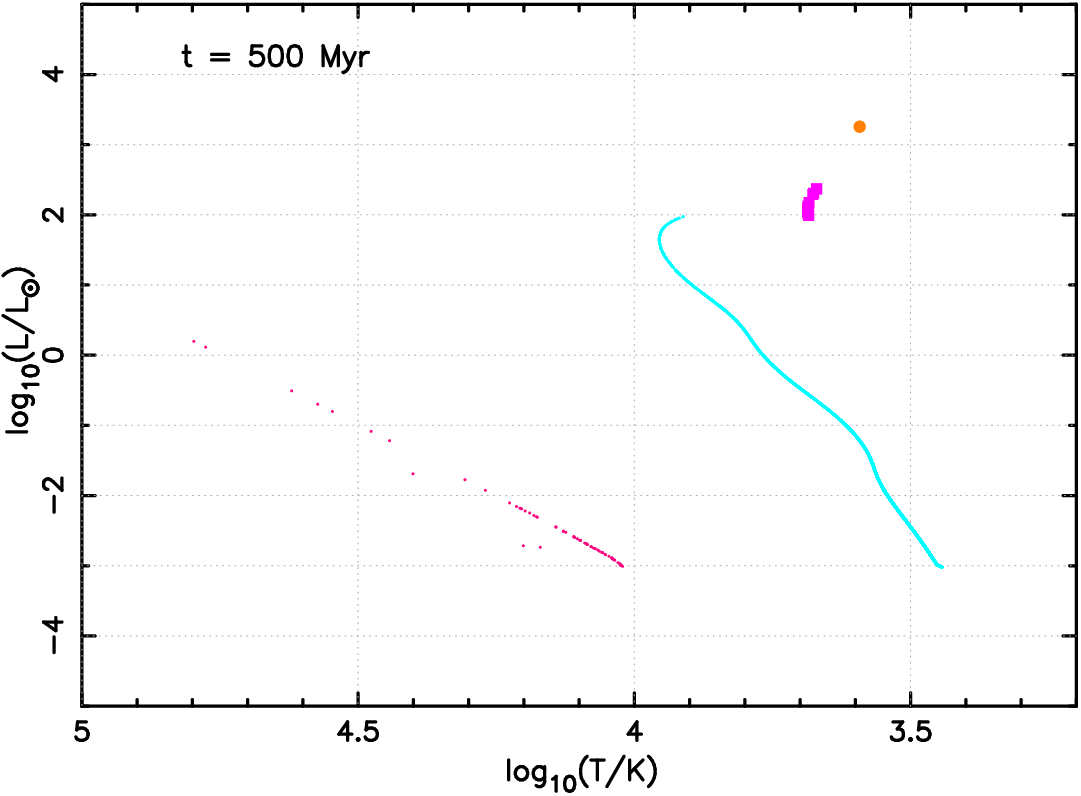}
\includegraphics[width=\columnwidth]{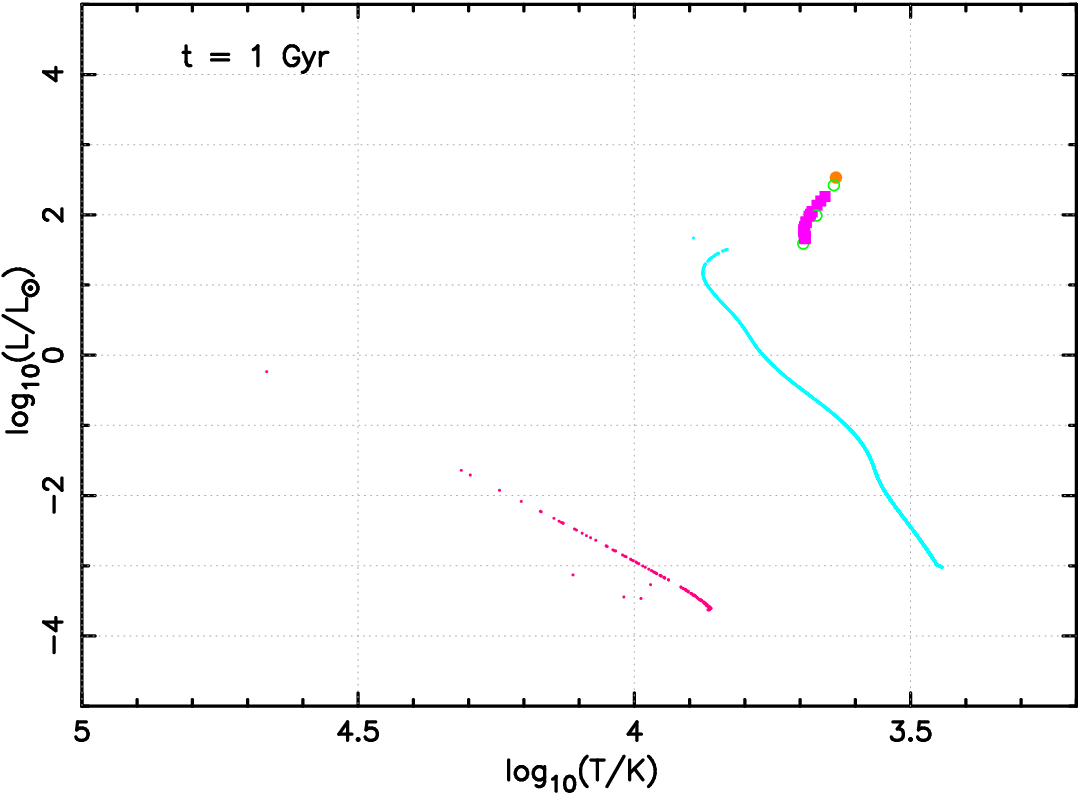}
\includegraphics[width=\columnwidth]{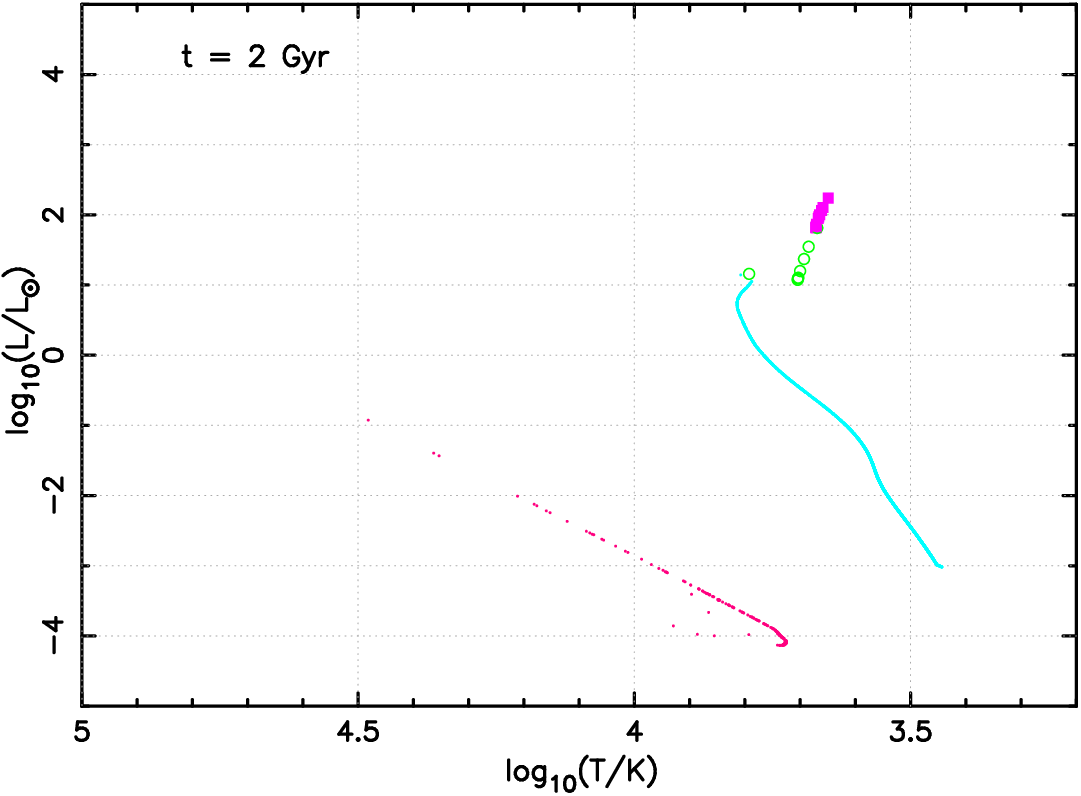}
\includegraphics[width=\columnwidth]{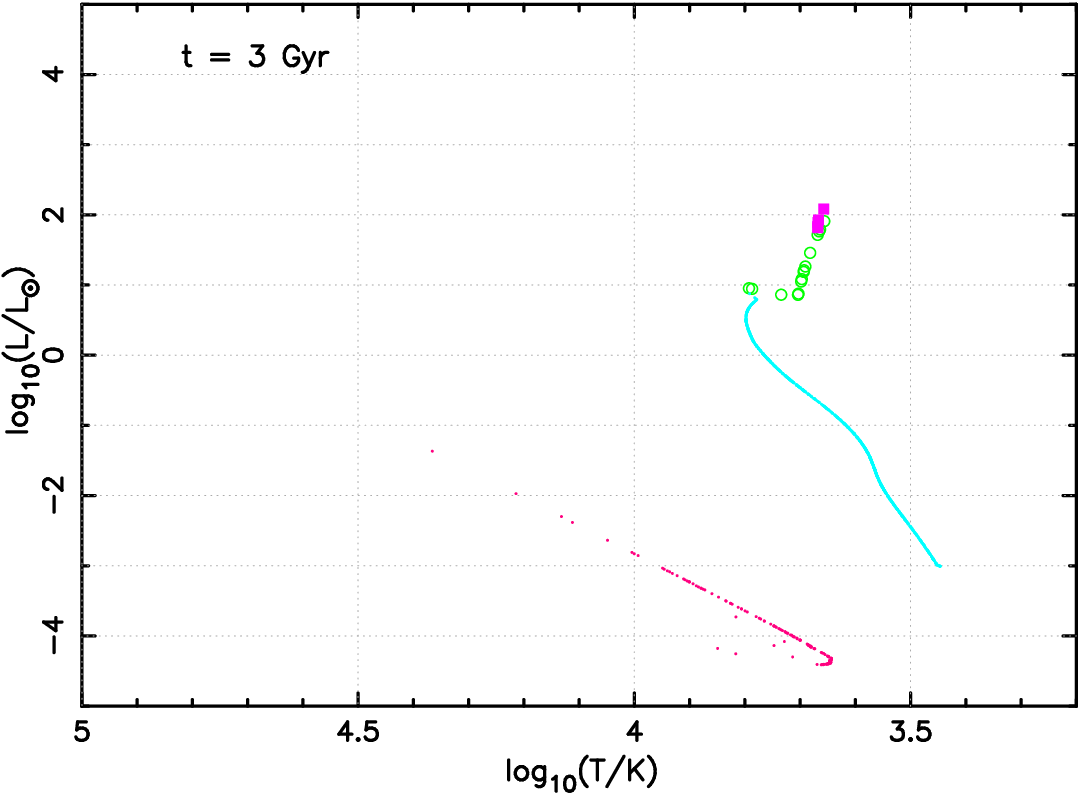}
\includegraphics[width=\columnwidth]{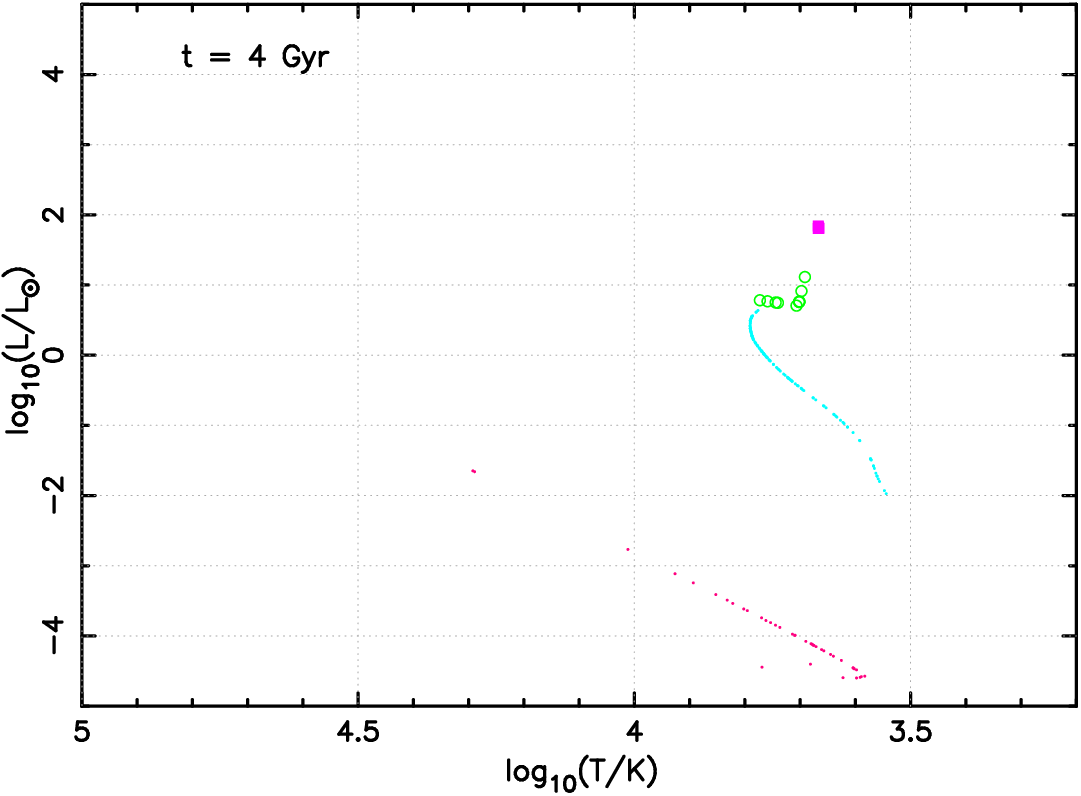}
\caption{As in Figure~\ref{fig:HR:STARS} for one of the \SSE~models.}
\label{fig:HR:SSE}
\end{center}
\end{figure*}

\begin{figure*}
\begin{center}
\includegraphics[width=\columnwidth]{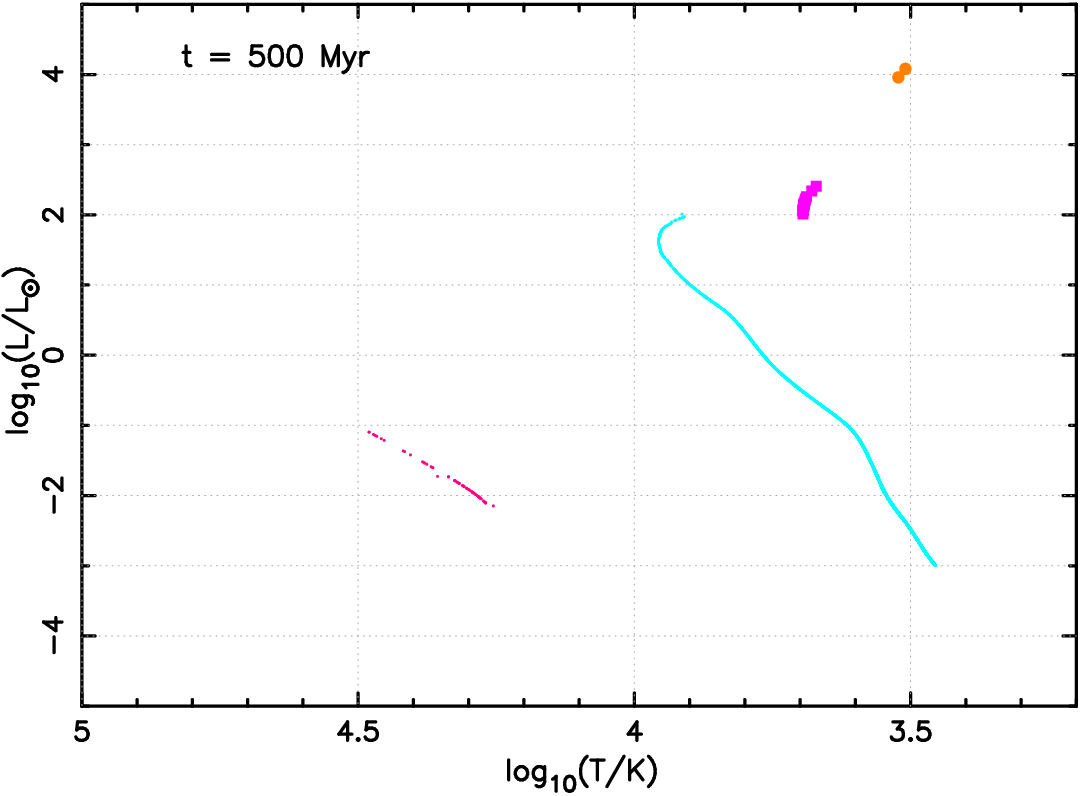}
\includegraphics[width=\columnwidth]{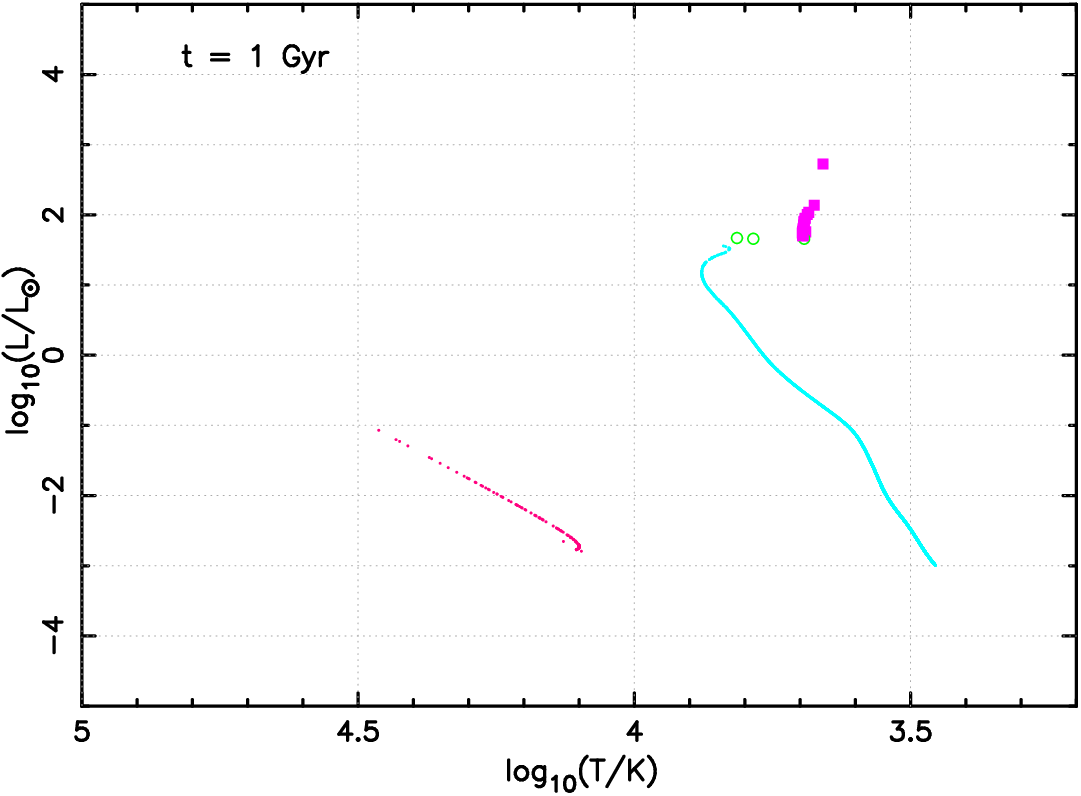}
\includegraphics[width=\columnwidth]{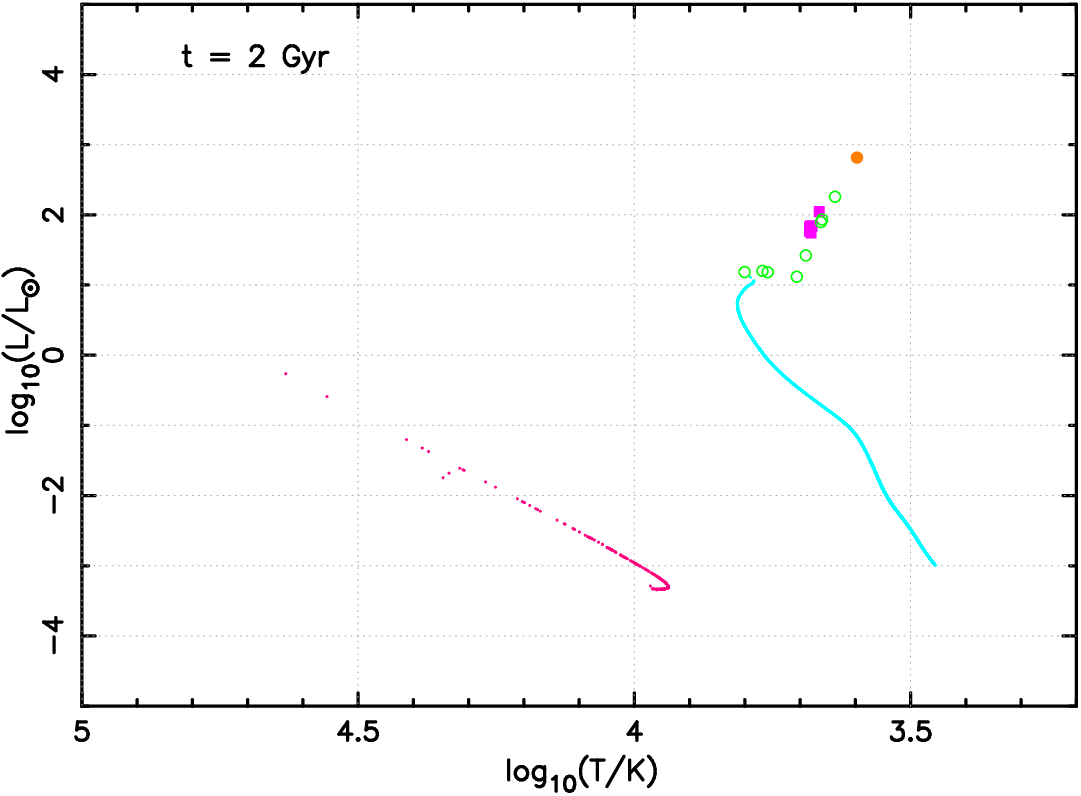}
\includegraphics[width=\columnwidth]{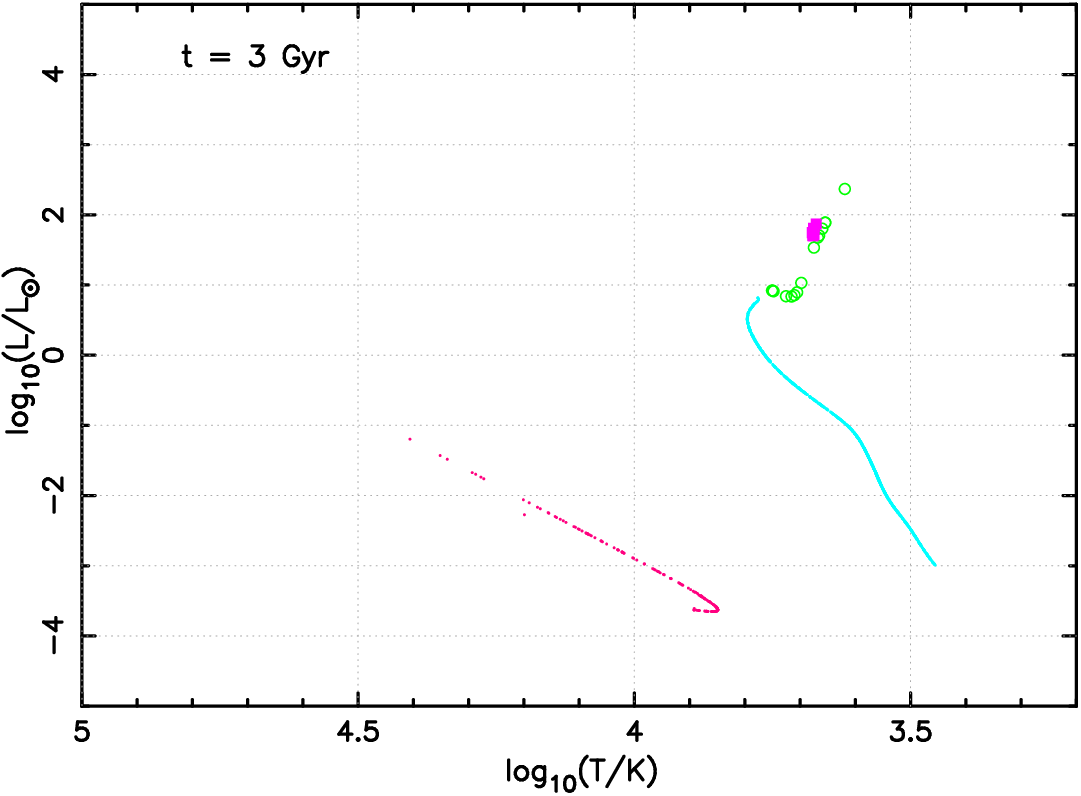}
\includegraphics[width=\columnwidth]{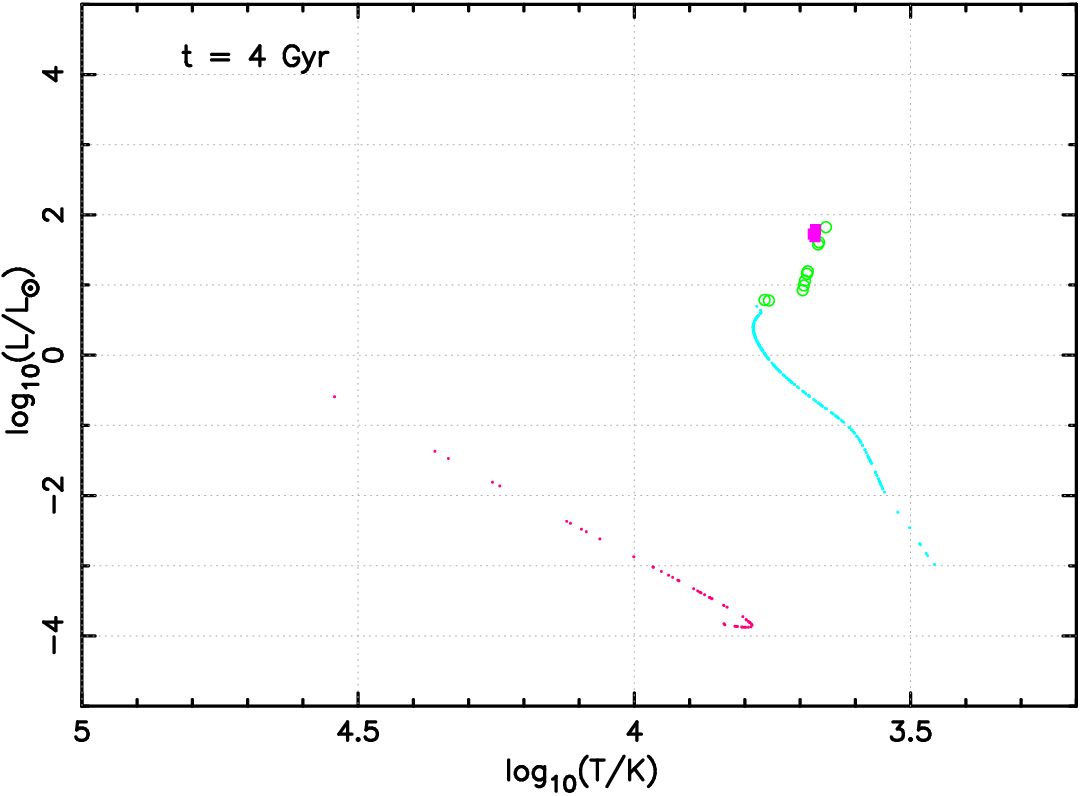}
\caption{As in Figure~\ref{fig:HR:STARS} for the \STARS~model with convective
overshooting.}
\label{fig:HR:ov}
\end{center}
\end{figure*}

The HR diagram of an actual cluster is relatively easily observed -- only
photometry and the distance to the cluster are required for absolute magnitudes.
Similarly the theoretical version, the luminosity-temperature diagram, is easily
generated as the temperature and luminosity are always calculated by a stellar
evolution code.  Comparison of the two requires bolometric correction via fits
to detailed models of stellar atmospheres.  However, even given the
uncertainties in this process, such diagrams provide a powerful and widely-used
tool for comparing models and observations of clusters.  Hence producing an
accurate HR diagram is an important function of the stellar evolution section of
a hybrid code.  Figure~\ref{fig:HR:STARS} contains HR diagrams for snapshots
from one of the \STARS~models without convective overshooting,
Figure~\ref{fig:HR:SSE} HR diagrams for one of the \SSE~models and
Figure~\ref{fig:HR:ov} diagrams for the \STARS~model computed with convective
overshooting.

\subsection{Stellar type fractions and mass}
\begin{figure*}
\begin{center}
\includegraphics[angle=270,width=0.66\columnwidth]{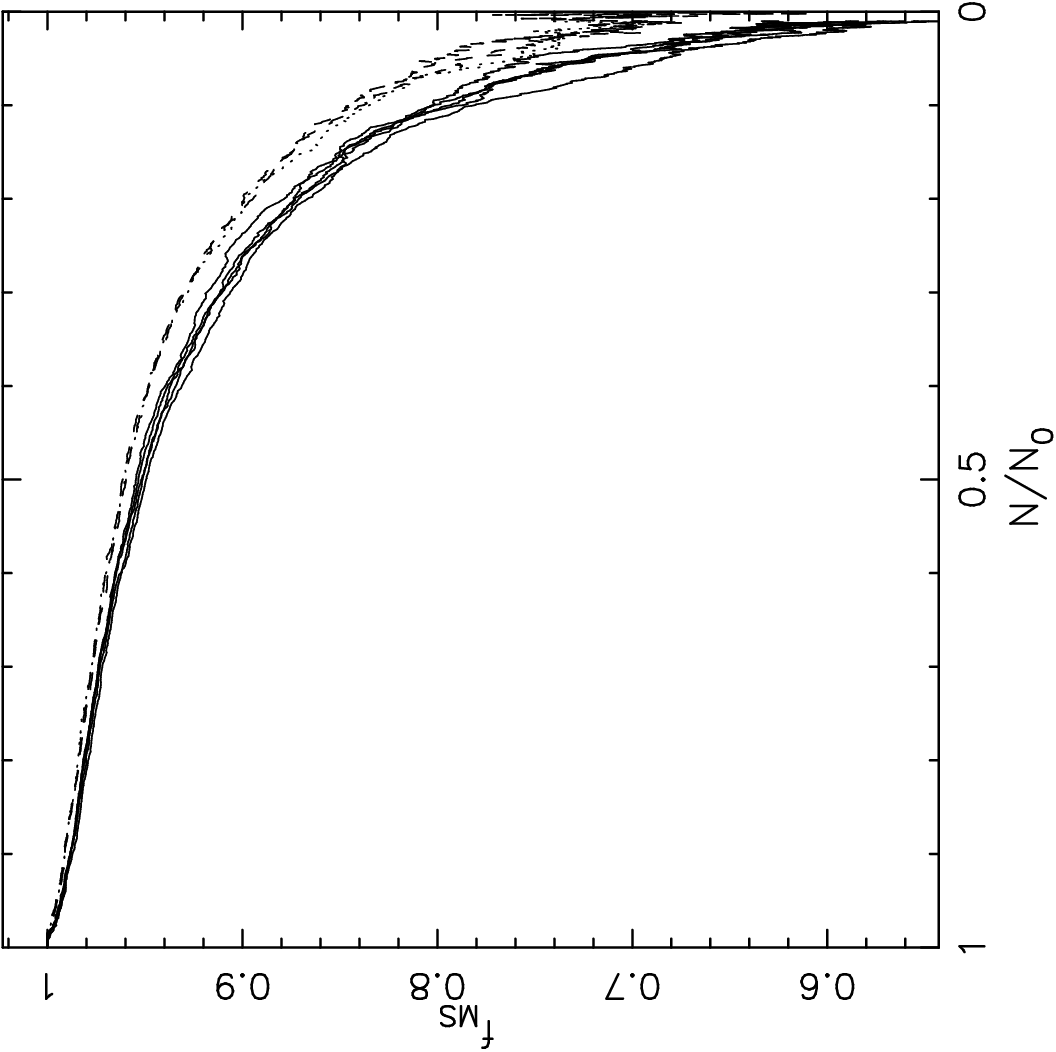}
\includegraphics[angle=270,width=0.66\columnwidth]{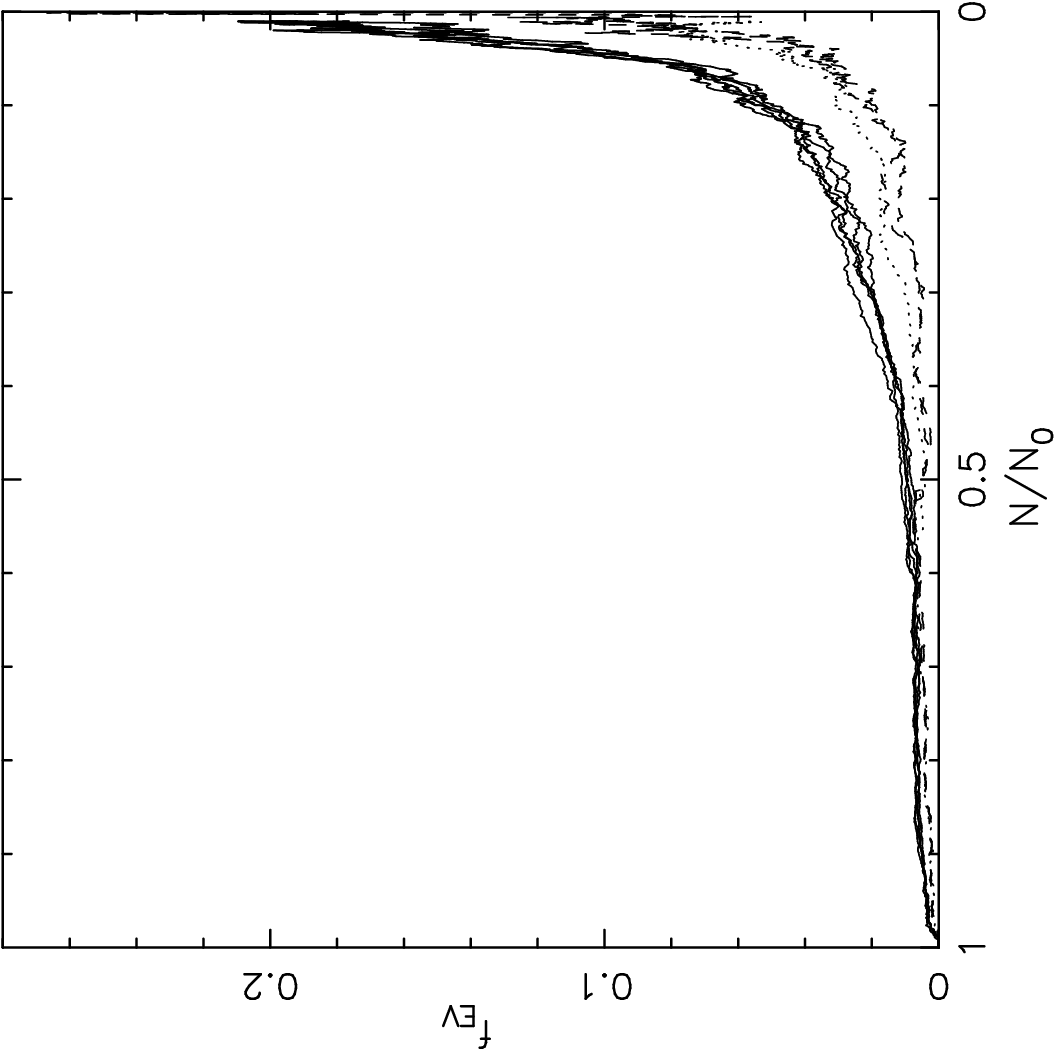}
\includegraphics[angle=270,width=0.66\columnwidth]{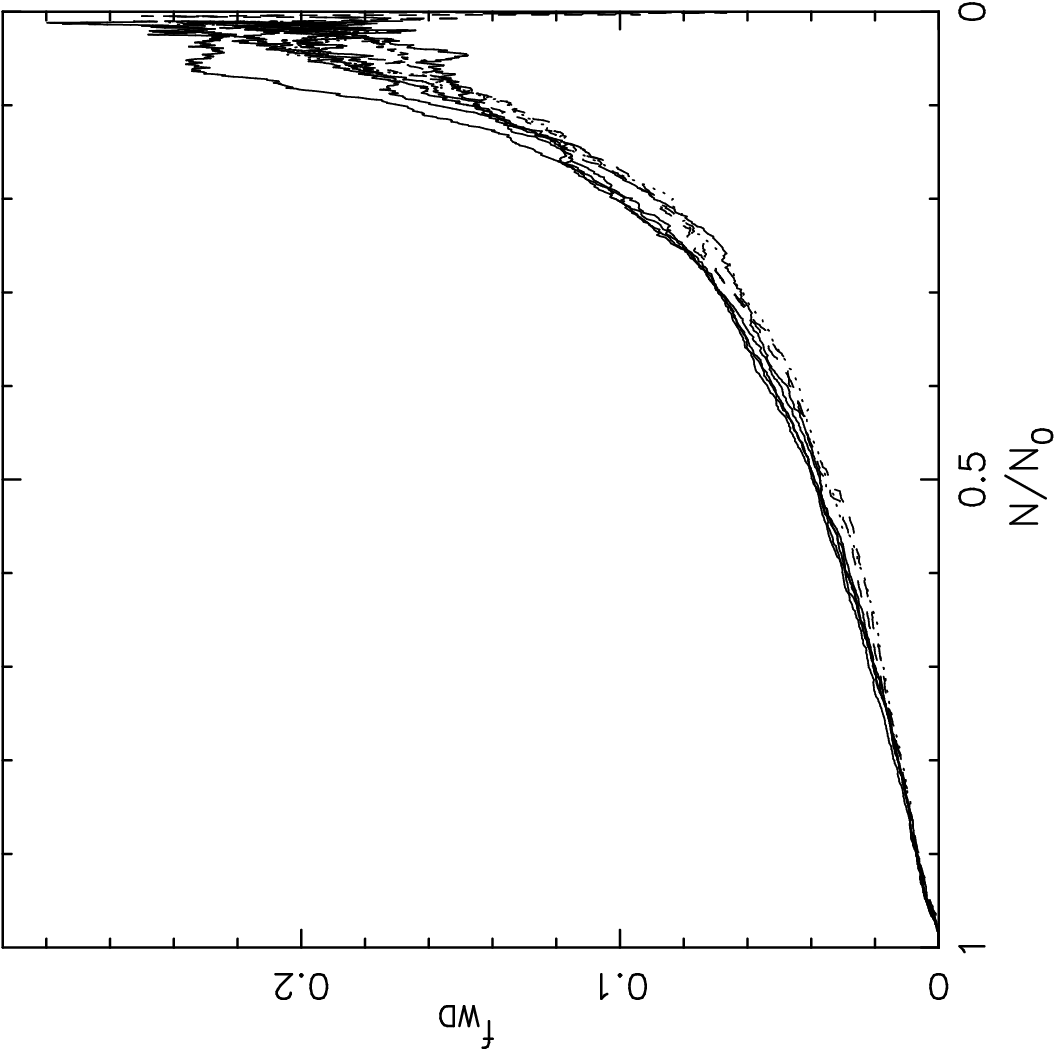}\\
\includegraphics[angle=270,width=0.66\columnwidth]{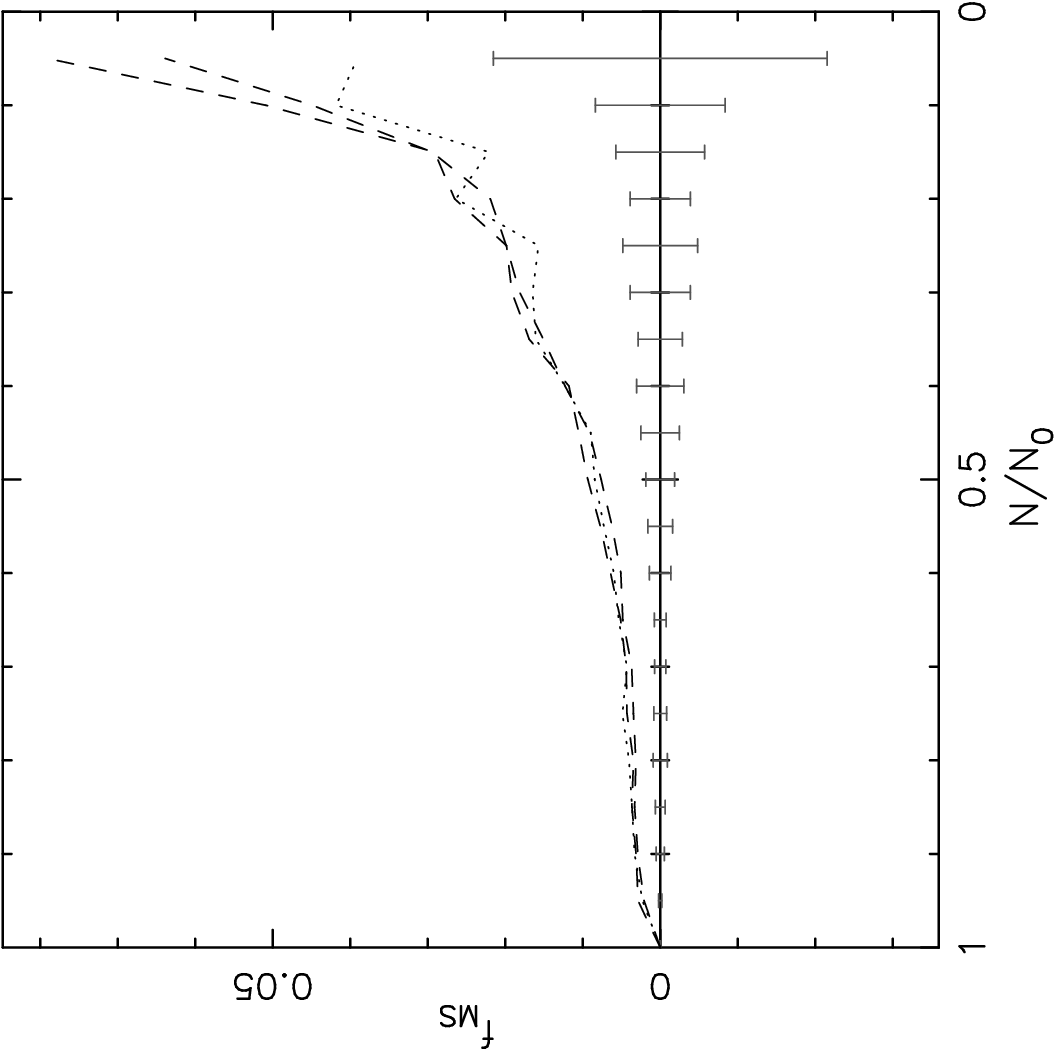}
\includegraphics[angle=270,width=0.66\columnwidth]{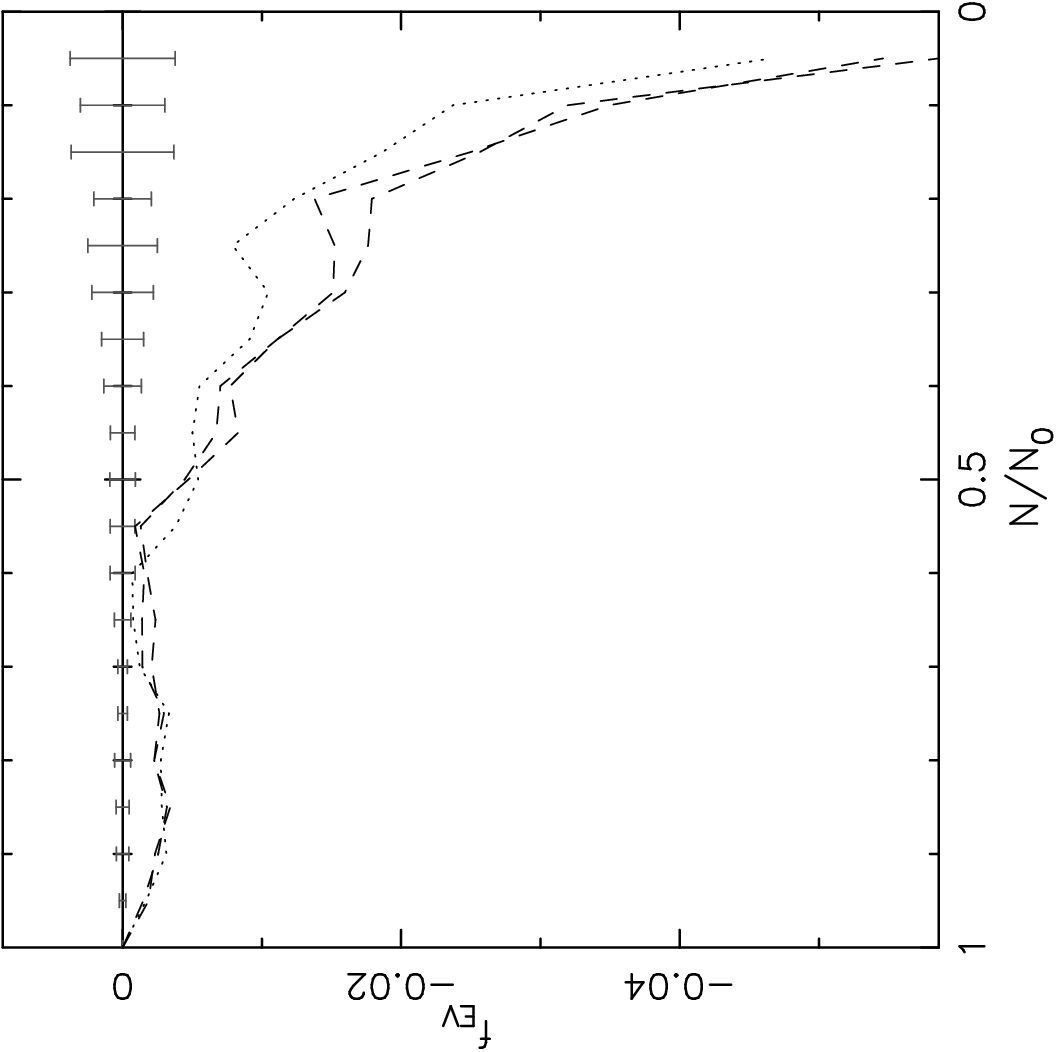}
\includegraphics[angle=270,width=0.66\columnwidth]{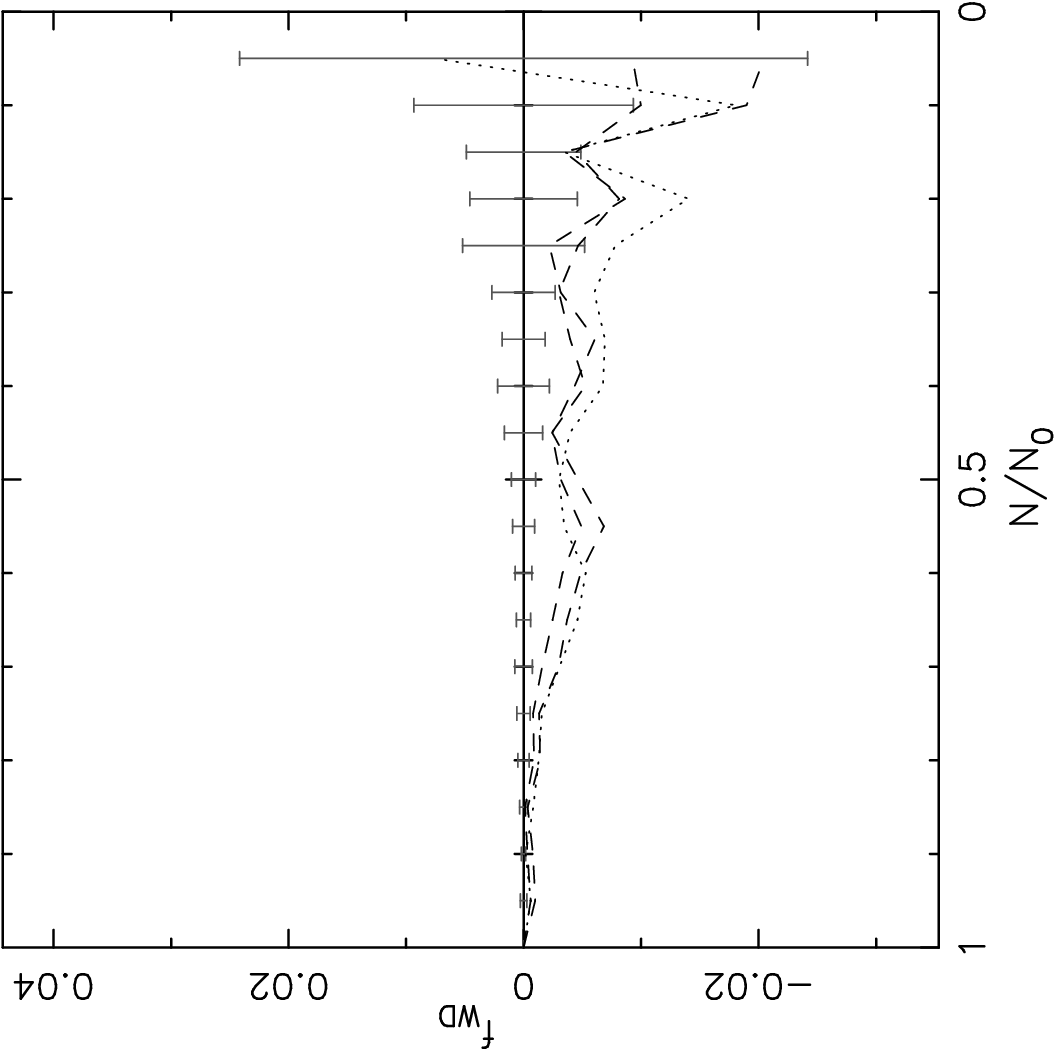}
\caption{The top left panel shows the fraction of stars that are on the main
sequence as a function of the fraction of stars remaining in the cluster.  The
solid lines are the \STARS~models without convective overshooting, the dashed
lines the \SSE~models and the dotted line the \STARS~model with convective
overshooting.  The bottom left panel shows the deviations of the latter two
values from the mean value of the five \STARS\ models run without convective
overshooting.  The grey error bars show the standard deviation of
these five models, the dashed lines the \SSE\ models and the dotted line the
\STARS\ model with convective overshooting.  The central panels show the same
measurements for evolved stars (giants and core helium burning stars) and the
right panels those for white dwarfs.}
\label{fig:types}
\end{center}
\end{figure*}

\begin{figure*}
\begin{center}
\includegraphics[angle=270,width=\columnwidth]{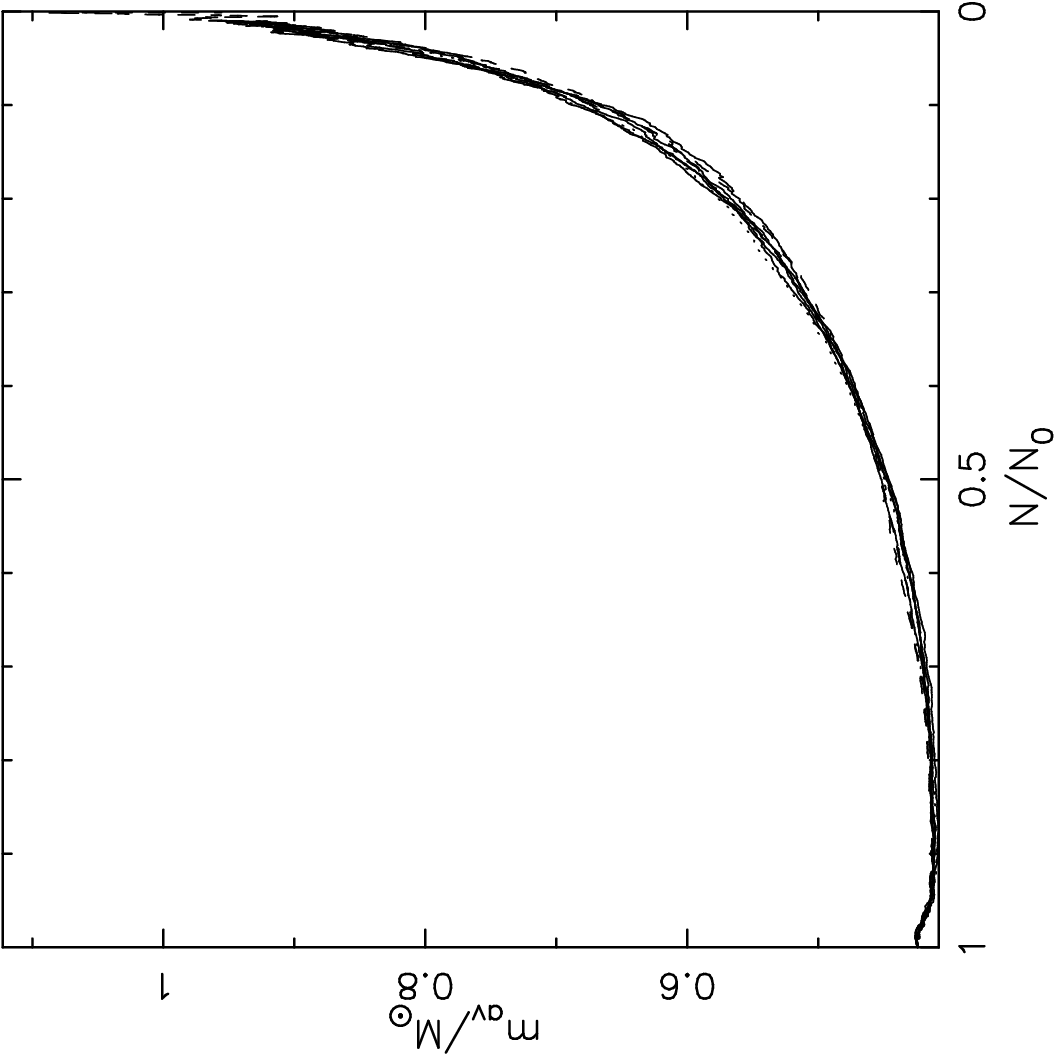}
\includegraphics[angle=270,width=\columnwidth]{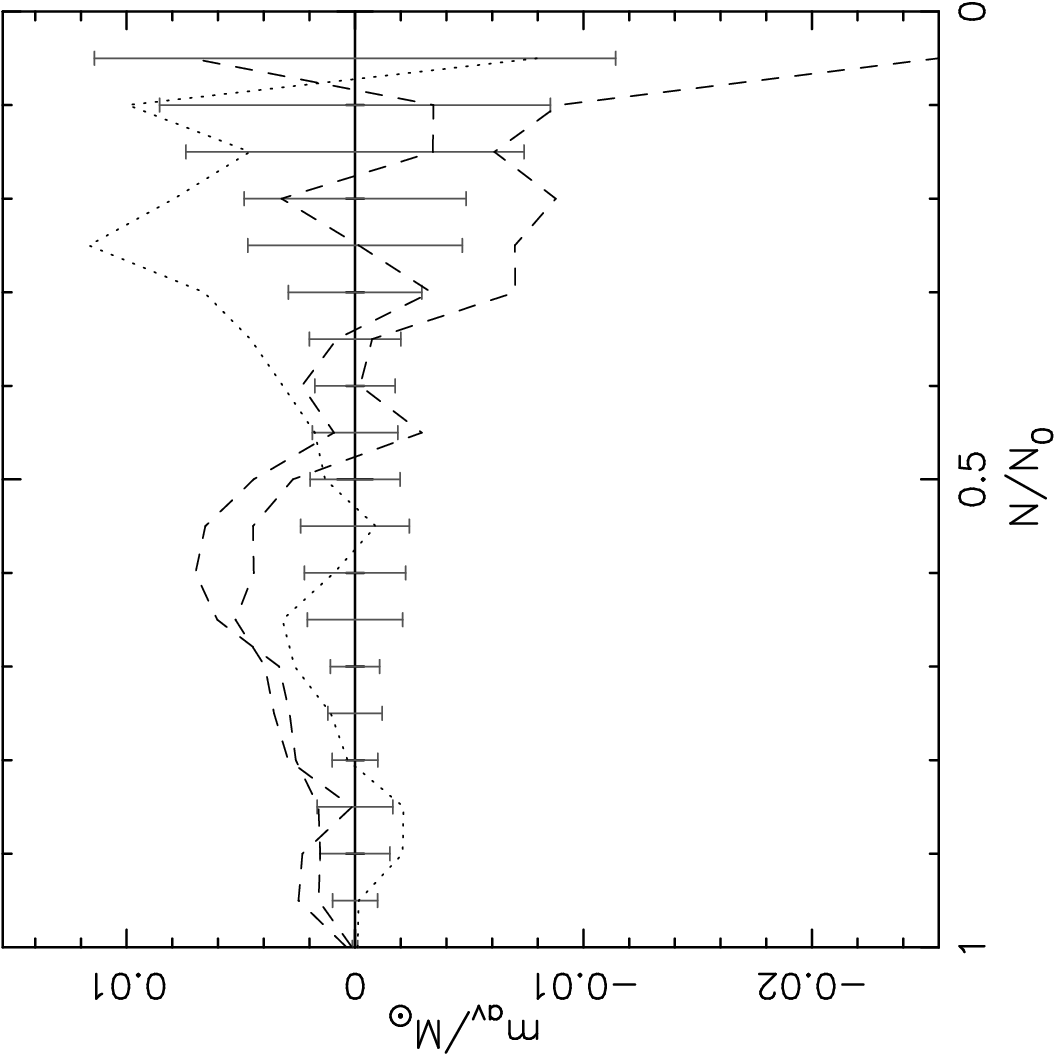}
\caption{The left panel shows the average stellar mass in the models plotted
against the fraction of stars remaining in the cluster.  The right panel shows
the deviations of the values from the mean of the five \STARS\ models calculated
without convective overshooting.  The solid lines are the \STARS\ models without
convective overshooting, the dashed lines the \SSE\ models and the dotted line
the \STARS\ model with convective overshooting.  The error bars represent the
standard deviation of the five \STARS\ models without convective overshooting.}
\label{fig:mav}
\end{center}
\end{figure*}

At the beginning of a model cluster's life its constituent stars are all on the
zero-age main sequence.  As time passes stars both evolve and are lost from the
cluster.  Both these processes affect the fraction of stars that are of a given
type and hence these properties depend on both dynamics and stellar evolution.
Figure~\ref{fig:types} contains diagrams showing the evolution of the various
stellar type fractions.  The same processes also affect the average stellar
mass, the evolution of which is shown in Figure~\ref{fig:mav}.  The conversion
of the fraction of stars remaining to time can be found from Figure~\ref{fig:t}.

\subsection{Time against number of stars}
\begin{figure*}
\begin{center}
\includegraphics[angle=270,width=\columnwidth]{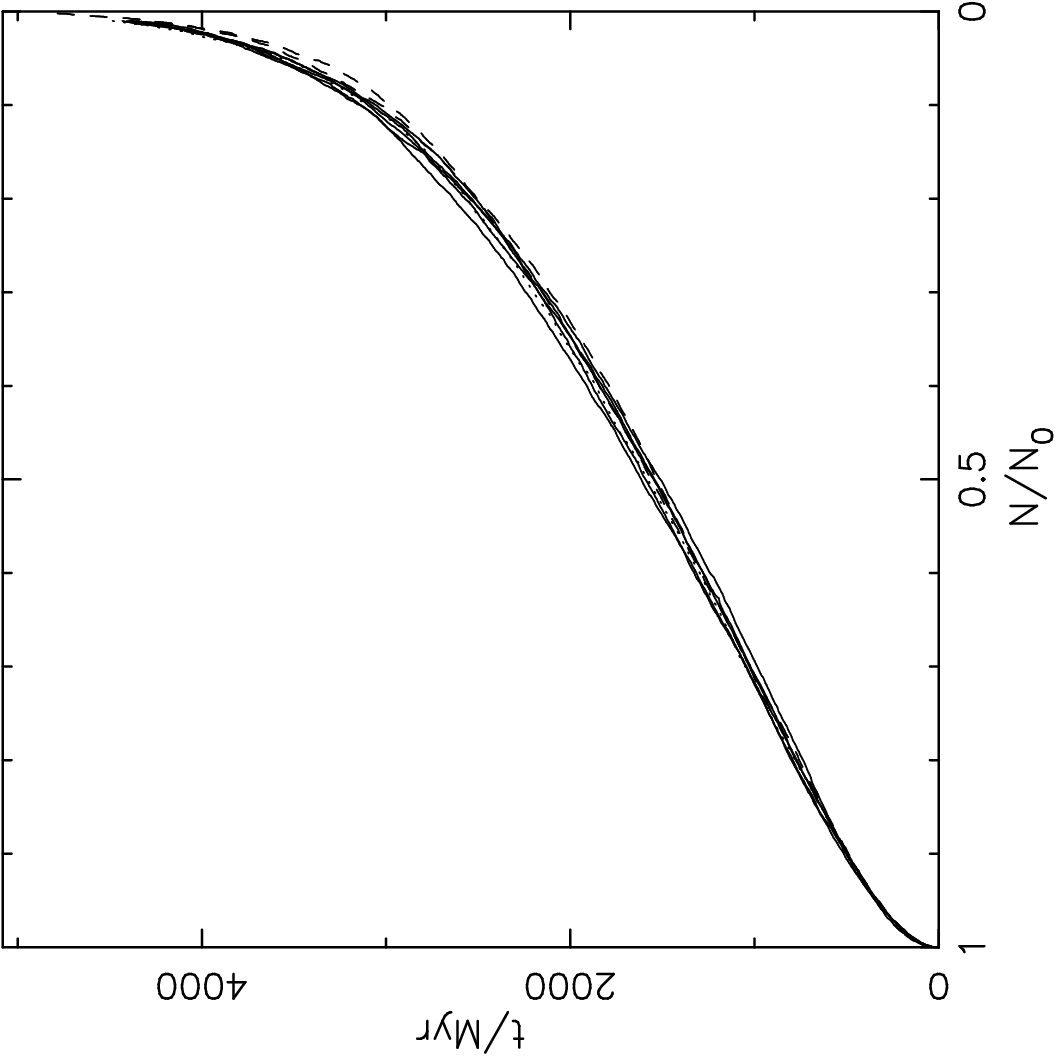}
\includegraphics[angle=270,width=\columnwidth]{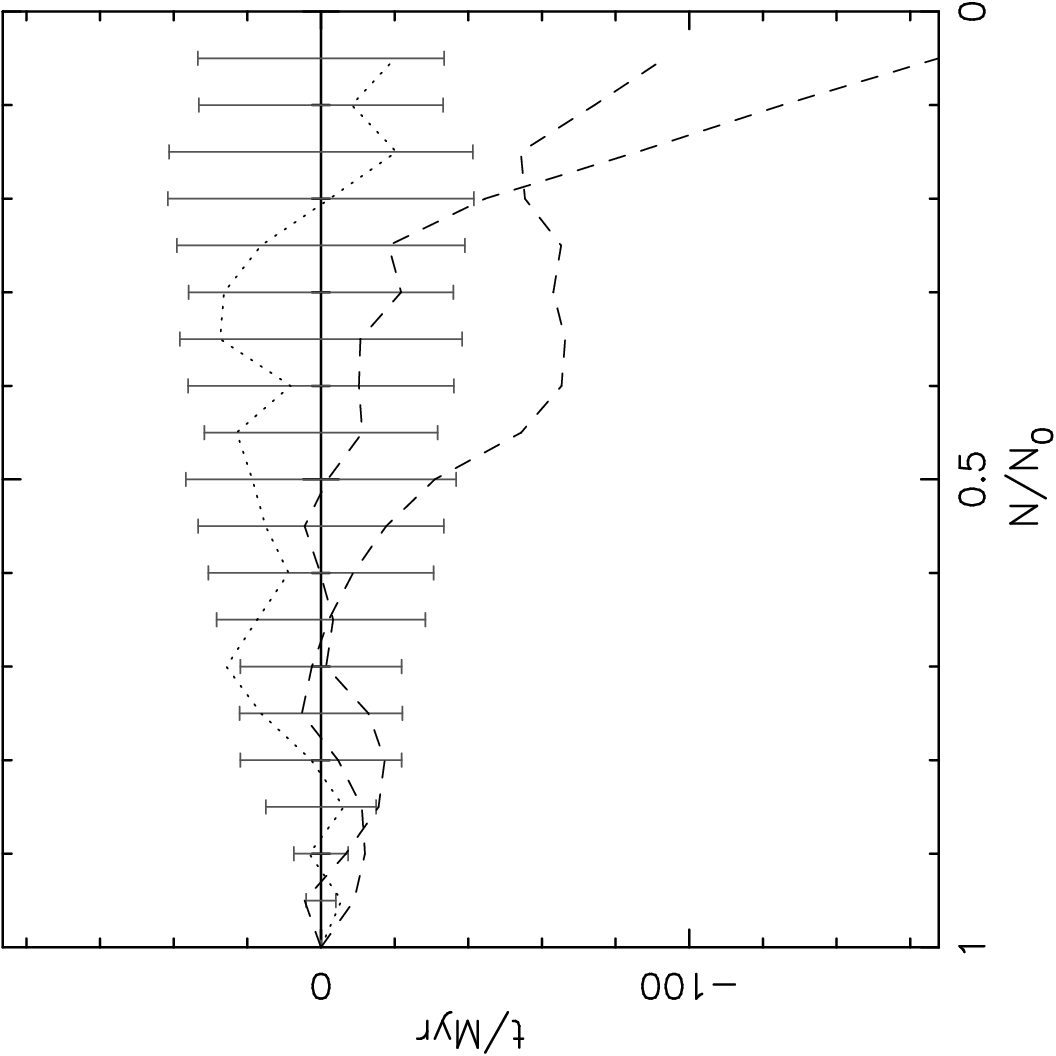}
\caption{As in Figure~\ref{fig:mav}, but for the physical time as a function
of the fraction of stars remaining in the cluster.}
\label{fig:t}
\end{center}
\end{figure*}

\begin{figure*}
\begin{center}
\includegraphics[angle=270,width=\columnwidth]{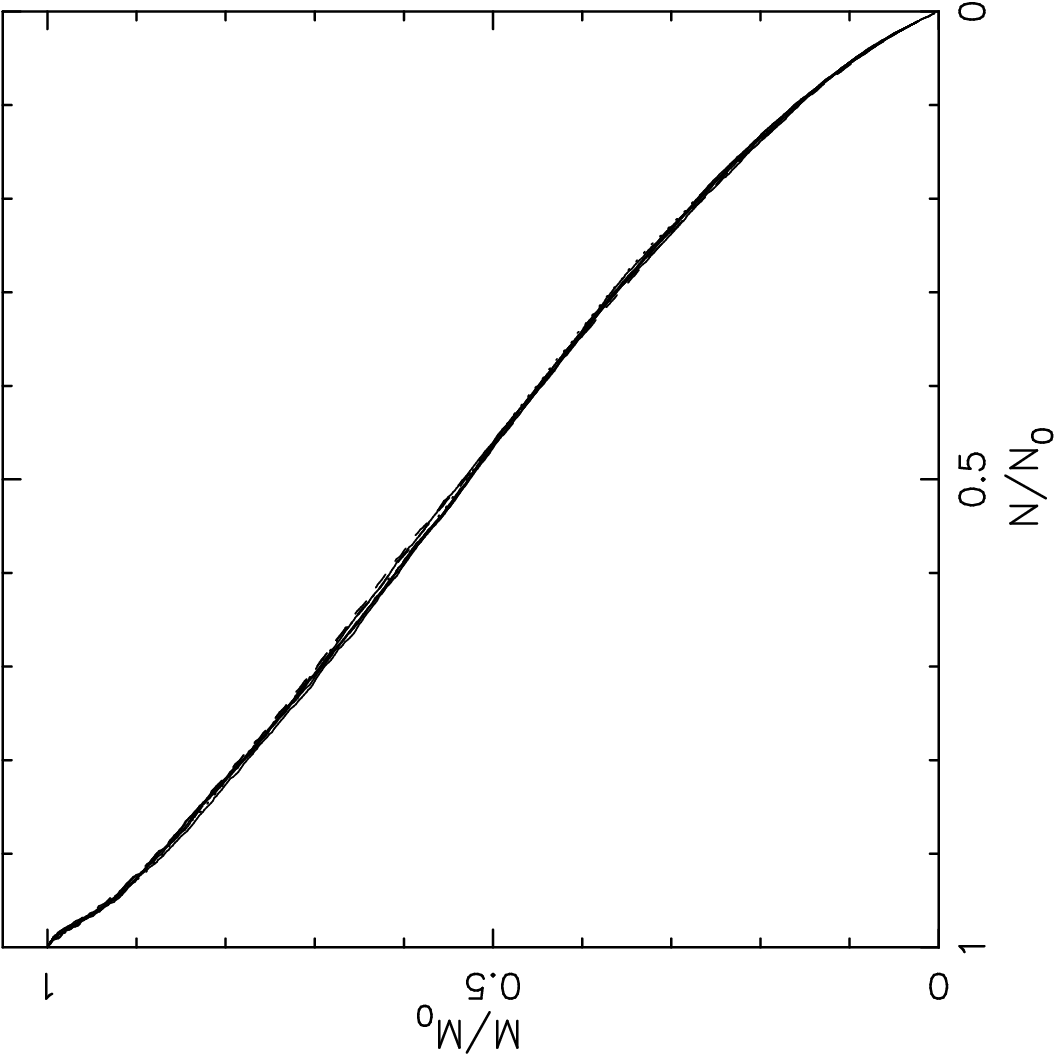}
\includegraphics[angle=270,width=\columnwidth]{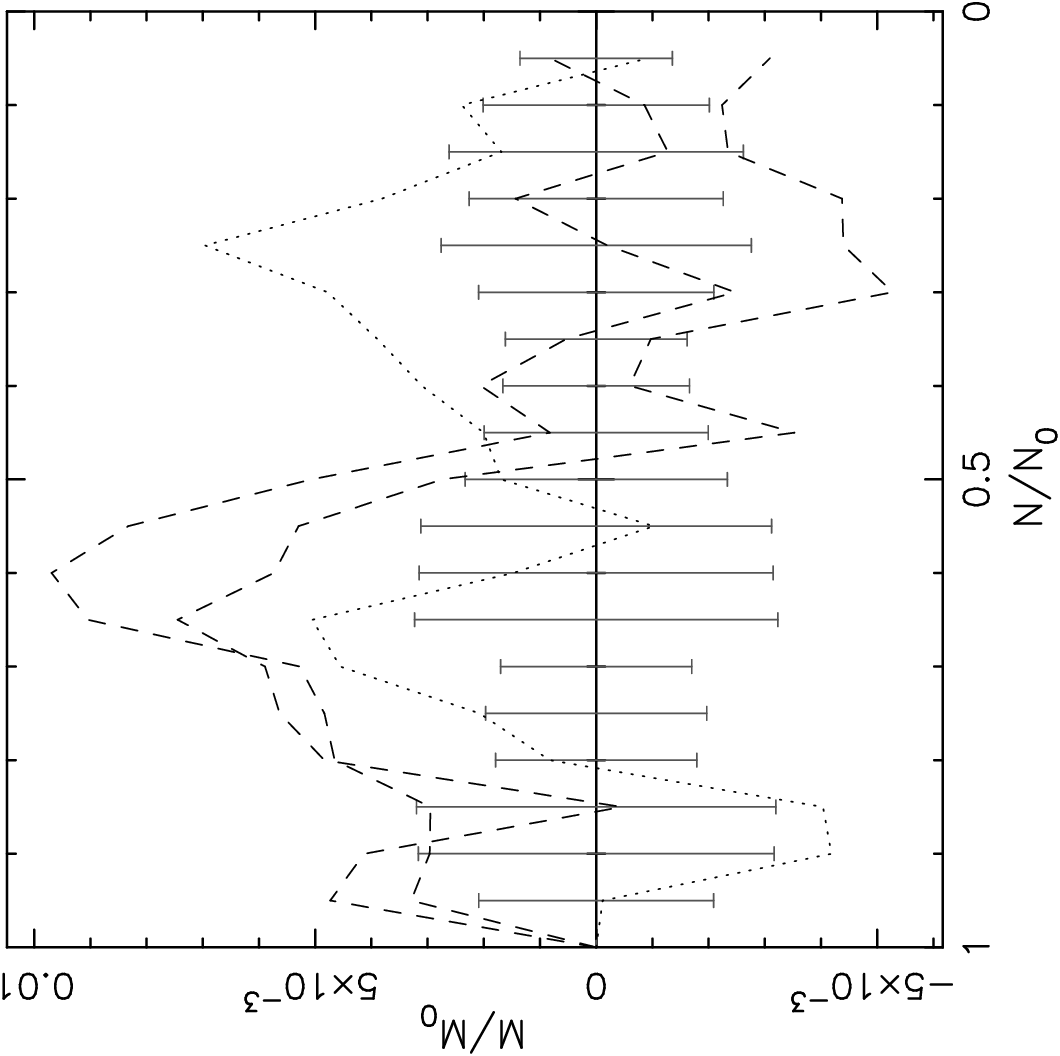}
\caption{As in Figure~\ref{fig:mav} but for the fraction of cluster mass
remaining as a function of the fraction of stars remaining.}
\label{fig:M}
\end{center}
\end{figure*}

Stars escape from a cluster by evaporation.  Few-body interactions exchange
energy between stars and an energy-gaining star can become unbound and escape
from the cluster.  This process is accelerated by the presence of the Galactic
tidal field.  Simultaneously the cluster loses mass because of evaporation,
stellar winds and supernovae.  This reduces the strength of the cluster
potential and hence increases the evaporation rate.  Thus the cluster mass and
the number of stars remaining are tracers of dynamical processes happening on a
smaller scale.  A plot of the time taken for the fraction of stars remaining to
fall to a given level is shown in Figure~\ref{fig:t} and a plot of the fraction
of the initial mass remaining in Figure~\ref{fig:M}.

\subsection{Core and half-mass radii}
\begin{figure*}
\begin{center}
\includegraphics[width=\columnwidth]{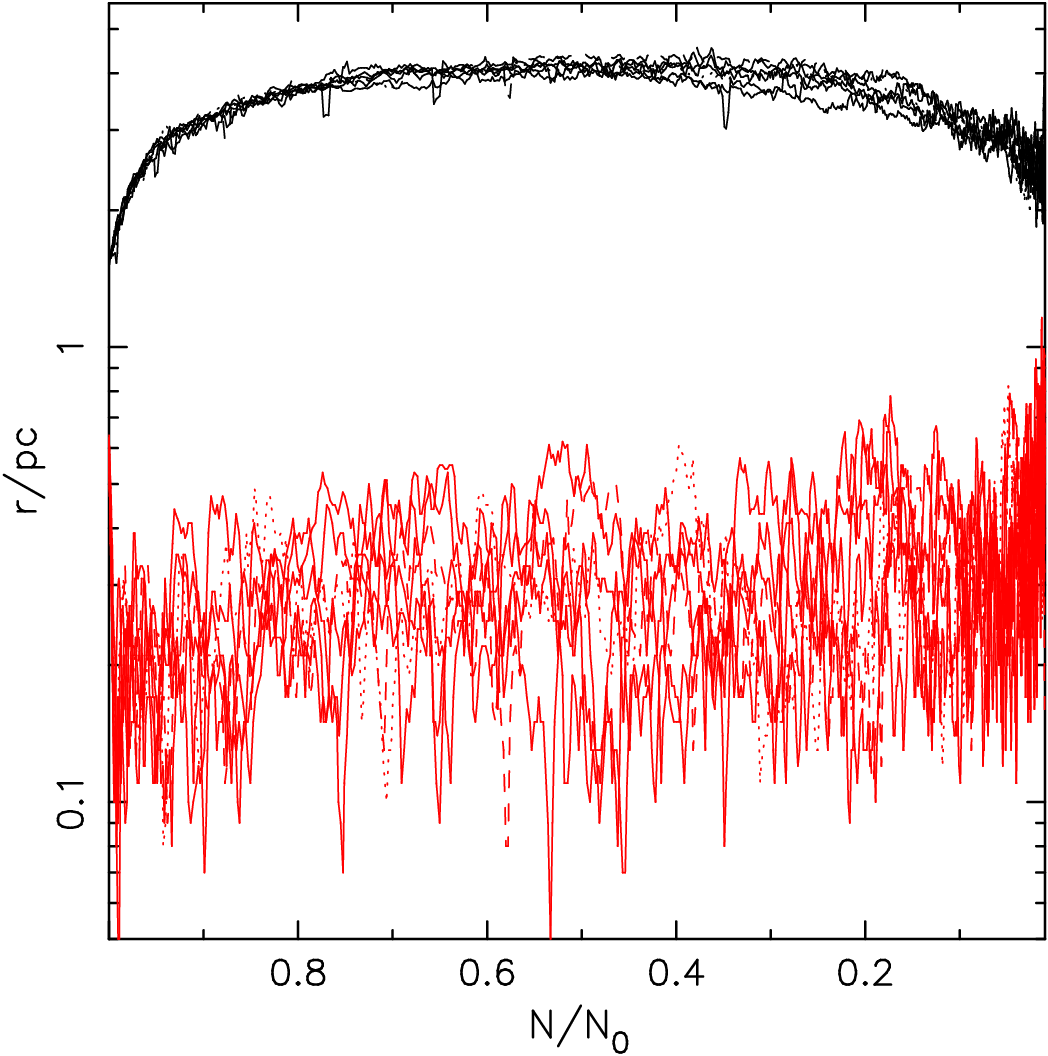}
\includegraphics[width=\columnwidth]{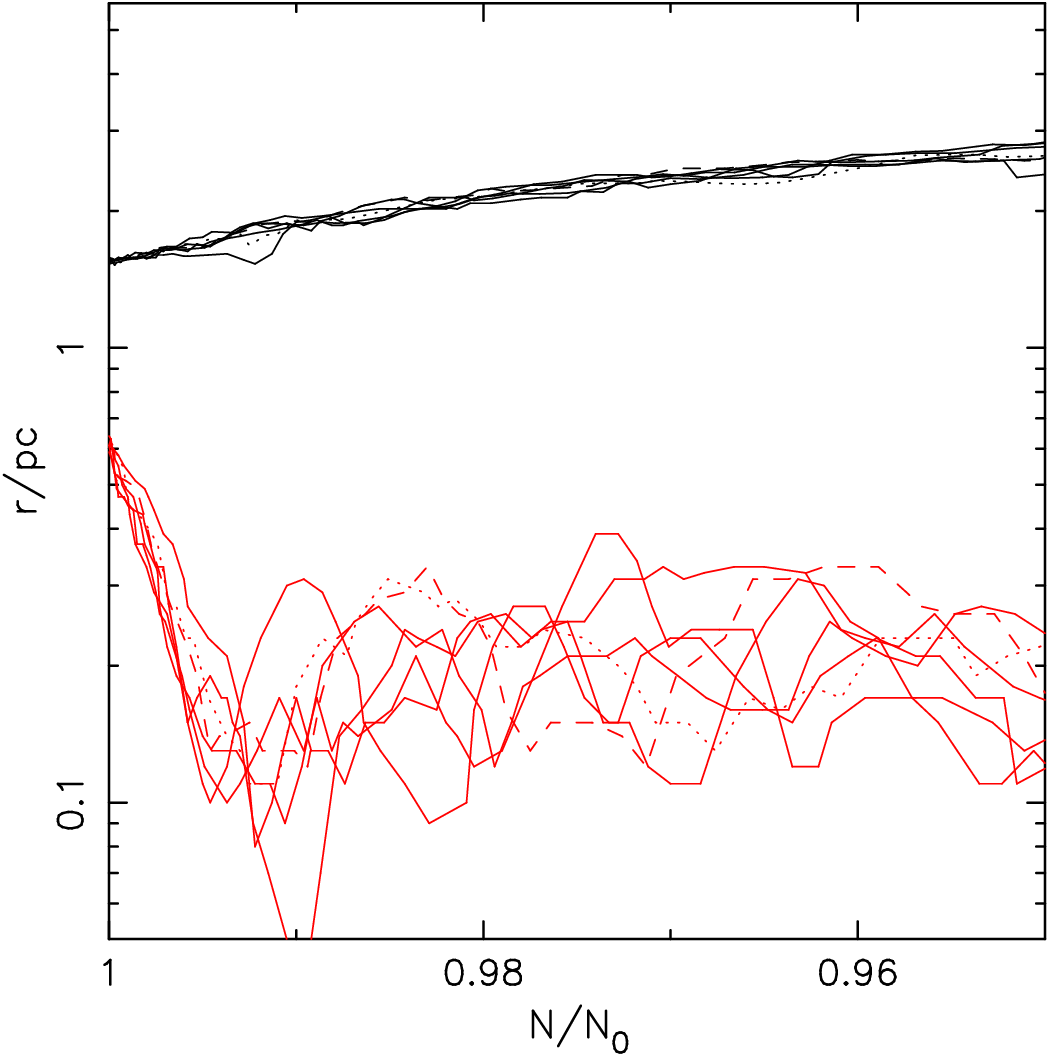}
\caption{The evolution of core radius (lower, grey line) and half-mass radius
(upper, black line) with fraction of stars remaining.  The values of the two
radii have been smoothed by averaging across three consecutive values.  The five
solid lines are the \STARS~models without convective overshooting, the dashed
line one of the \SSE~models and the dotted line the \STARS~model with convective
overshooting.  The right panel shows just the first part of the evolution to
facilitate observation of core collapse.}
\label{fig:radii}
\end{center}
\end{figure*}

The core is the most important part of a stellar cluster.  In the
core densities are highest, so the closest encounters between stars take
place.  The crossing time of the core is smaller than that of the whole cluster,
so catastrophic collapse owing to the gravothermal instability takes place there
first.  The core radius is the standard indicator of how large and how dense the
core is, whilst the half-mass radius allows us to measure the behaviour of the
outer parts of the cluster.  A plot of the evolution of the core and half-mass
radii with time is presented in Figure~\ref{fig:radii}.

\section{Comparison and discussion}
\label{sect:discuss}
The results show that the behaviour of the different codes is very similar, with
some small differences which we discuss individually.

\subsection{Stellar measures}
Comparing Figures~\ref{fig:HR:STARS}~and~\ref{fig:HR:SSE} several differences
can be seen.  The Hertzsprung gap in the \STARS\ model is much more populated --
this is particularly evident at 2\,Gyr.  The turnoff mass is lower in the
\STARS\ models because their main-sequence lifetimes are shorter.
Finally the minimum white dwarf temperatures in the \STARS\ models are higher
suggesting that they are cooling more slowly.  This occurs even though the
turnoff mass is lower, which implies that the white dwarfs formed earlier.

The first two differences are caused by the presence of convective overshooting
in the models from which \SSE\ was constructed.  Convective overshooting -- the
mixing of material outside the boundaries of convective regions in stars -- is a
possible explanation for the extra mixing over and above normal convective
mixing that it thought to take place in stars~\citep{1973ApJ...184..191S}.
Amongst the several effects it has on evolution, described
in~\citet{1997MNRAS.285..696S}, it extends the main-sequence lifetime and reduces
the time spent crossing the Hertzsprung gap.  The set of HR diagrams in
Figure~\ref{fig:HR:ov}, for a cluster run using \STARS\ with convective
overshooting, show that the first two differences largely disappear.  Although
there are still one or two more stars in the Hertzsprung gap the difference is
not substantial.  The differences in the white dwarf cooling track are caused by
the over-simplified physics included in the version of \SSE\ that we were using.
This has been remedied in more recent versions.

The overall trend in the evolution of the stellar type distributions, as shown
in Figure~\ref{fig:types}, is that the fraction of main-sequence stars decreases
as the cluster evolves whereas the fractions of white dwarfs and evolved stars
increase.  The decline in main-sequence star numbers is caused by stars evolving
off the main sequence and the preferential evaporation of low-mass stars which
are predominantly on the main sequence.  The number of white dwarfs is larger
than the number of evolved stars other than at very early times because the
evolved stars subsequently evolve further into white dwarfs.  On the other hand
as the cluster approaches dissolution the fraction of evolved stars increases
much more rapidly, because these are now some of the heaviest stars in the
cluster and hence the last to be lost.

Again it can be seen from the plots of residuals that the inclusion of
convective overshooting has a significant effect on cluster evolution.  The
fraction of main-sequence stars in the two models with convective overshooting
(the \SSE~models and the \STARS\ model with convective overshooting) decline
less quickly than in the \STARS~model without convective overshooting,
particularly at the expense of evolved stars.  For stars of intermediate mass
overshooting enlarges the convective core causing more hydrogen to be burnt and
hence increasing the main-sequence lifetime.  The \SSE\ models and the \STARS\
model with convective overshooting are seen to agree within the intrinsic
variability of the problem given by the error bars.

\subsection{Dynamical properties}
Comparing the results for \SSE~and \STARS~in Figure~\ref{fig:t} we can see that
for most of the models the total number of stars and mass of stars in the
cluster are in good agreement between the two codes.  The \SSE~models have
fewer stars towards the end of the evolutionary sequence (after about half the
stars have been lost from the simulations).  To explain this difference it is
useful to consider first the average stellar mass.

We can see in Figure~\ref{fig:mav} that the average stellar mass behaves
similarly across all the models.  Initially it declines slightly owing to mass
loss from the most massive stars and then increases following the preferential
evaporation of low-mass stars.  From the plot of the residuals we see
that until about half the stars in the cluster have been lost the \SSE{} models
have higher average masses.  This is because the inclusion of convective
overshooting increases the main-sequence lifetimes of intermediate-mass stars.
This in turn increases the turnoff mass and hence the mean stellar mass.
However as the cluster evolves the stars at the turnoff have smaller and smaller
convective cores and so the difference in lifetimes is less pronounced.  The
residuals of the average mass for the \STARS{} model with convective
overshooting shows a similar trend but displaced downward slightly, as the mean
mass drops more sharply at the beginning of the simulation.  This is presumably
because there are, by chance, a larger number of the highest mass stars in the
simulation.  These effects are also clear in the plot of the total mass
remaining in the cluster (Figure~\ref{fig:M}).

The effect of this enhanced number of more massive stars appears to be to
increase the rate of ejection of stars from the cluster.  This explains the
number of stars being lower in the \SSE{} models from the point where half the
stars have evaporated onwards.  However, one should be aware of reading too much
into the results; few of the differences are significant beyond two standard
deviations.

From the plot of core and half-mass radii (Figure~\ref{fig:radii}) it is hard to
glean much interesting information.  Looking at the half-mass radius we can see
that the cluster expands initially, owing to mass loss from evolving stars, then
remains at roughly constant size for most of its lifetime and shrinks towards
the end as it dissolves.  The core radius declines sharply at the start of the
simulation, undergoes core collapse just prior to 200\,Myr and behaves
erratically thereafter.  No significant deviation between the different sets of
stellar models can be identified.

\section{Conclusions}

We would expect, {\it a priori}, that the introduction of full stellar
evolution would have a limited effect on cluster simulations containing only
single stars because the synthetic stellar evolution fits are generally found to
be good for single stars.  These results show that this is indeed the case.
Once convective overshooting has been included in the models they fit the
results obtained with \SSE\ to within the intrinsic variability of the
calculations.

One of the benefits of a live stellar evolution code is the increased
flexibility that it brings.  In order to change the value of the convective
overshoot parameter, the mixing length parameter for convective mixing, the
initial elemental abundances, or any other aspect of the physics package, inside
\STARS~we merely a change one or two parameters.  This contrasts with a
synthetic code where a whole new grid of models must be computed and fits made
to them.  Hence a live code, whilst not offering much improvement in accuracy
for single stars, does provide additional flexibility.  The new code allows us
to produce, for instance, stellar isochrones and luminosity functions that
incorporate the effect of dynamics, whilst also allowing us to vary aspects of
the stellar physics such as the mixing and initial chemical compositions.

To make full use of detailed stellar evolution models we need to extend the code
described here.  The problems with numerical non-convergence on the late
AGB/post-AGB need to be solved for stars of mass greater than $4\,\Mo$ so that
we can simulate the whole range of masses of single stars.  The next step will
then be to add the effects of binary interactions.  Interactions are very
significant for the dynamical evolution of clusters and binaries are responsible
for the formation of many different types of stellar exotica, including blue
stragglers, X-ray binaries, millisecond pulsars, etc.  For a full description of
cluster evolution, this important (and often difficult) area of evolution needs
to be modelled properly.  Such processes as Roche Lobe overflow and common
envelope evolution must be included.  Work on developing the code further to
this end is underway.

\section{Acknowledgements}

The authors would like to thank Sverre Aarseth for the use of the \NB6 $N$-body
code and much patient assistance during its adoption and use; without his help
this work would have not been possible.  RPC is grateful to PPARC for a
scholarship and to the Australian Research Council Discovery Project for support
under grant DP0663447.  CAT is grateful to Prof. John Lattanzio for funding his
sabbatical leave in Australia and to Churchill College Cambridge for a
fellowship.  JRH is grateful to the Australian Research Council for a
fellowship.

\bibliography{paper1}

\end{document}